\documentclass[apj,twocolumn,iop]{emulateapj}
\usepackage[hyperfootnotes=false,colorlinks,citecolor=blue,linkcolor=blue,urlcolor=blue]{hyperref}
\newcommand{\Mo}{M_\odot}

\begin{document}

\title{On Physical Scales of Dark Matter Halos}

\author{Marcel Zemp}
\affil{Kavli Institute for Astronomy and Astrophysics, Peking University, Yi He Yuan Lu 5, Hai Dian Qu, 100871 Beijing, China}
\email{mzemp@pku.edu.cn}

\begin{abstract}
It is common practice to describe formal size and mass scales of dark matter halos as spherical overdensities with respect to an evolving density threshold.
Here, we critically investigate the evolutionary effects of several such commonly used definitions and compare them to the halo evolution within fixed physical scales as well as to the evolution of other intrinsic physical properties of dark matter halos.
It is shown that, in general, the traditional way of characterizing sizes and masses of halos dramatically overpredicts the degree of evolution in the last 10 Gyr, especially for low-mass halos.
This pseudo-evolution leads to the illusion of growth even though there are no major changes within fixed physical scales.
Such formal size definitions also serve as proxies for the virialized region of a halo in the literature.
In general, those spherical overdensity scales do not coincide with the virialized region.
A physically more precise nomenclature would be to simply characterize them by their very definition instead of calling such formal size and mass definitions 'virial'.
In general, we find a discrepancy between the evolution of the underlying physical structure of dark matter halos seen in cosmological structure formation simulations and pseudo-evolving formal virial quantities.
We question the importance of the role of formal virial quantities currently ubiquitously used in descriptions, models and relations that involve properties of dark matter structures.
Concepts and relations based on pseudo-evolving formal virial quantities do not properly reflect the actual evolution of dark matter halos and lead to an inaccurate picture of the physical evolution of our universe.
\end{abstract}

\keywords{galaxies: halos --- galaxies: structure --- galaxies: kinematics and dynamics --- galaxies: evolution --- dark matter --- methods: numerical}

\section{Introduction}

Within the currently favored cosmological framework where the energy content is dominated by dark energy and cold dark matter  and space is flat ($\Lambda$CDM), peaks of the initial matter density field become gravitationally unstable, collapse, then virialize and finally form what we call dark matter halos.
Embedded within those dark matter halos is the baryonic matter that forms the galaxies, which we can observe.
For a recent, detailed and pedagogical review on the theory of cosmological structure formation see \cite{2012arXiv1208.5931K} and references therein.
Therefore, for a comprehensive understanding of the universe, it is of fundamental importance to understand how structures form in cosmology and how dark matter halos and their enclosed galaxies evolve with cosmic time.

In the simplest conceivable scenario, one can study the collapse of a spherically symmetric density perturbation in an otherwise homogeneous universe \citep{1972ApJ...176....1G}.
This is called the spherical collapse model and we review its essentials in Section \ref{sec:spherical_overdensity}.
When combined with the spectrum of (known) perturbations, this results in an analytic and elegant model for structure formation in the universe.
This approach was pioneered in the seminal work of \cite{1974ApJ...187..425P} and led to fundamental concepts of how we describe the size and mass of dark matter halos and their abundance (i.e. mass function), that are used to describe the properties of our universe.

The increase of computer power over the last few decades has enabled the possibility to simulate the formation of structure in the universe with ever better and more detailed models.
This has led to a good understanding of the large-scale structure of the dark matter \citep[e.g.][]{2005Natur.435..629S,2009MNRAS.398.1150B,2009A&A...497..335T,2011JKAS...44...217,2011ApJ...740..102K}, as well as the detailed small-scale and phase-space structure of individual dark matter halos \citep[e.g.][]{2007ApJ...667..859D,2008Natur.454..735D,2008ApJ...680L..25D,2008MNRAS.391.1685S,2009MNRAS.398L..21S,2009MNRAS.395..797V,2009MNRAS.394..641Z}.
Nowadays, the major challenge for computational structure formation simulations is to model the complex physics of star formation and feedback in a robust fashion.

In order to describe the properties and distribution of dark matter halos in such simulations, it is common practice to use concepts that are rooted in the simplistic analytic models from the early days of the era of cosmological structure formation.
Concretely, inspired by the spherical collapse model, the sizes and masses of dark matter halos are defined with respect to a density threshold that in general evolves with cosmic time.
Such definitions lead to pseudo-evolution: an artificial growth of the size and mass of a dark matter halo without much happening within fixed physical scales \citep[see also appendix \ref{sec:mechanism} for an illustration of the pseudo-evolution mechanism]{2007ApJ...667..859D,2008MNRAS.389..385C,2013ApJ...766...25D}.
In this paper, we expand on earlier studies and shed further light on the issue from various, previously unexplored angles.
We also discuss potential solutions and more physical descriptions.

The structure of this paper is as follows.
In Section \ref{sec:spherical_overdensity}, we review the ideas and models that have inspired dark matter halo size/mass definitions based on spherical overdensities.
In Section \ref{sec:property_evolution}, we study the evolution of various properties of dark matter halos in a cosmological structure formation simulation over the last 10 Gyr.
This is then followed by a detailed discussion in Section \ref{sec:discussion}, where we discuss the implications on models, descriptions and relations within the realm of structure and galaxy formation.
We then conclude and summarize in Section \ref{sec:summary}.

\section{Characteristic scales based on spherical overdensities}\label{sec:spherical_overdensity}

The simple picture of the formation of halos in the universe within the framework of the spherical collapse model served as an inspiration for definitions of sizes and masses of halos commonly used.
Classically (see for example \cite{2008gady.book.....B}, section 9.2.1, or \cite{2012arXiv1208.5931K}, section 2.3.1), this is derived for an Einstein-de Sitter universe (i.e. $\Omega_\mathrm{M} = 1$, $\Omega_{\Lambda} = 0$, $\Omega_\mathrm{K} = 0$, $\rho_\mathrm{mean} = \rho_\mathrm{crit}$ at all times) where a spherical region is overdense with respect to the rest of the universe.
The matter is assumed to be cold (i.e. no shell crossing) and to have no angular momentum (i.e. only radial motions).
Therefore, the dynamics of the overdensity is solely determined by the total interior mass.
This overdensity will initially expand, but this expansion will slow down due to gravity and will come to a halt at the turnaround radius at a time $t_\mathrm{ta}$.
At this turnaround point, the density contrast of the sphere with respect to the mean matter density of the universe will be $\rho_\mathrm{sphere}(t_\mathrm{ta})/\rho_\mathrm{mean}(t_\mathrm{ta}) = 9 \pi^2/16 \approx 5.552$.
Theoretically, the oversimplified spherical collapse model would reach a singularity at $t_\mathrm{vir}=2 t_\mathrm{ta}$ but in reality the collapsing dark matter will virialize and settle into an equilibrium configuration which we call a halo.
Energy conservation and the virial theorem then imply that the size of the sphere is $r_\mathrm{vir} = r(t_\mathrm{ta})/2$.
Since for the mean matter density in an Einstein-de Sitter universe $\rho_\mathrm{mean} \propto t^{-2}$, the mean matter density drops by a factor 4 and the density within the sphere increases by a factor of 8.
Thus we get for the density contrast at virialization
\begin{equation}
  \Delta_\mathrm{vir} \equiv \frac{\rho_\mathrm{sphere}(t_\mathrm{vir})}{\rho_\mathrm{mean}(t_\mathrm{vir})} = 18 \pi^2 \approx 177.7~,
\end{equation}
i.e. the average density within the virialized region of size $r_\mathrm{vir}$ is given by
\begin{equation}
  \rho_\mathrm{vir} = 18 \pi^2 \rho_\mathrm{mean}
\end{equation}
and the mass is given by
\begin{equation}
  M_\mathrm{vir} = (4 \pi/3) r_\mathrm{vir}^{3} \rho_\mathrm{vir} \, .
\end{equation}

The classic derivation for the Einstein-de Sitter universe can be generalized for an open universe (i.e. $\Omega_\mathrm{M} < 1$, $\Omega_{\Lambda} = 0$, $\Omega_\mathrm{K} > 0$) or a $\Lambda$CDM universe (i.e. $\Omega_\mathrm{M} + \Omega_{\Lambda} = 1$, $\Omega_\mathrm{K}=0$) \citep{1993MNRAS.262..627L,1996MNRAS.282..263E,1998ApJ...503..569E,1998ApJ...495...80B}.
For the more relevant case of a $\Lambda$CDM universe, there are two common formulas that are used.
\cite{1998ApJ...503..569E} fit their calculations with
\begin{equation}\label{eq:E98}
  \rho_\mathrm{vir} = 178 \, \Omega_\mathrm{M}^{0.45} \rho_\mathrm{crit}
\end{equation}
whereas \cite{1998ApJ...495...80B} find the following fit
\begin{equation}\label{eq:BN98}
  \rho_\mathrm{vir} = (18 \pi^2 - 82 q - 39 q^2) \rho_\mathrm{crit}
\end{equation}
where
\begin{equation}
  q \equiv \frac{\Omega_{\Lambda,0}}{\Omega_\mathrm{M,0} a^{-3} + \Omega_{\Lambda,0}}
\end{equation}
and $a$ is the scale factor.

These definitions of size and mass of a halo inspired by the spherical collapse model can be put in a more general form
\begin{equation}
  \rho_\mathrm{SO} \equiv \frac{M_\mathrm{SO}}{(4 \pi/3) r_\mathrm{SO}^{3}} = \Delta \, \rho_\mathrm{ref}
\end{equation}
so that the halo is defined by the sphere where the average spherical overdensity (SO) reaches a specified threshold.
In general, $\Delta$ can be constant or a function of time and the reference density $\rho_\mathrm{ref}$ is either chosen to be the mean matter density of the universe $\rho_\mathrm{mean}$ (often also denoted as the background density) or the critical density of the universe $\rho_\mathrm{crit}$, which are both functions of time as well.

\begin{figure}
  \begin{center}
    \includegraphics[width=\columnwidth]{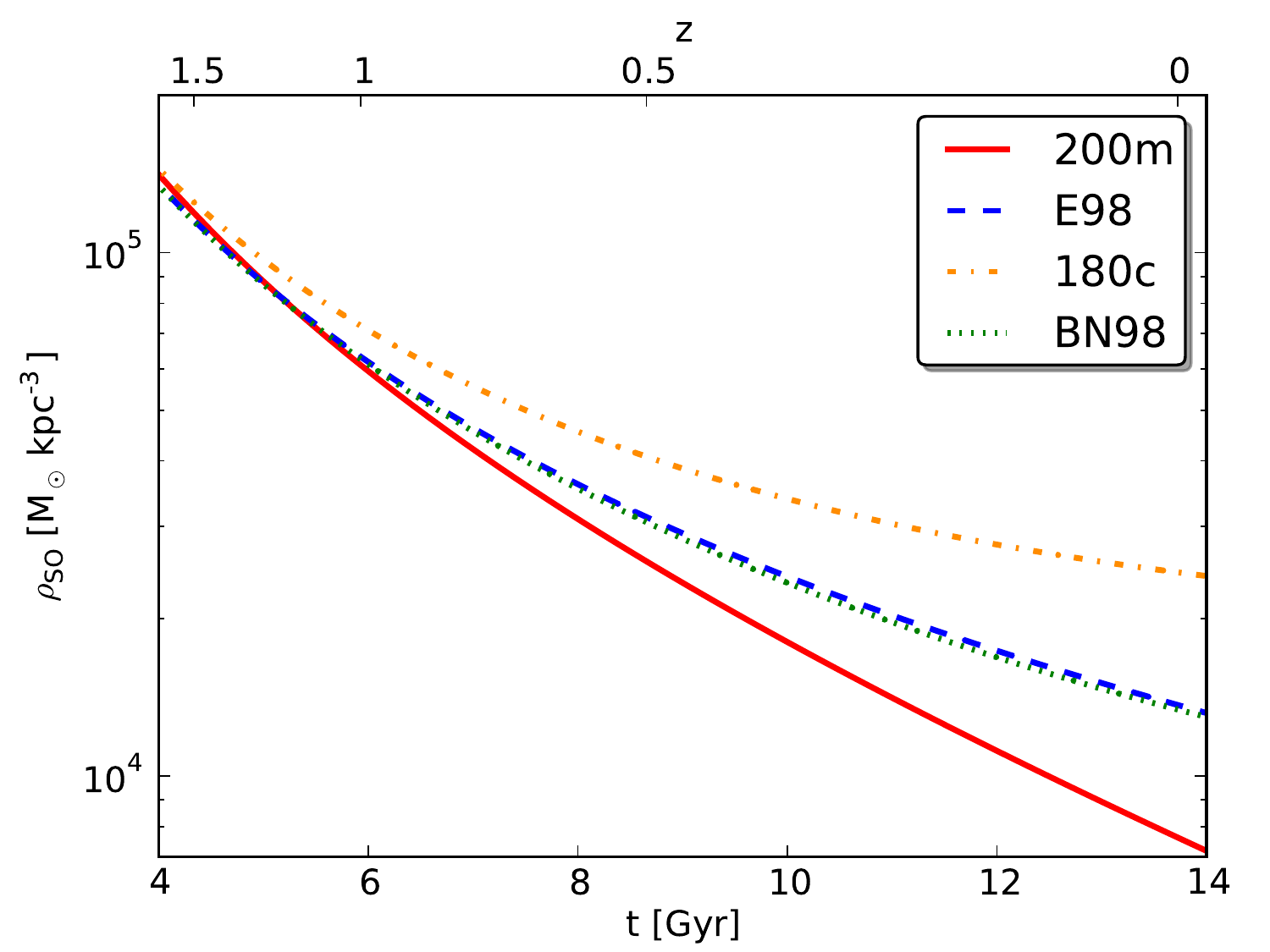}
    \caption{Evolution of 4 commonly used definitions for the spherical overdensity threshold $\rho_\mathrm{SO}$ as function of cosmic time since the big bang (bottom axis) respectively redshift (top axis) in a $\Lambda$CDM universe.
    It is evident that all of those 4 definitions evolve quite a bit over the shown range of approximately the last 10 Gyr.
    At high redshift $\Omega_\mathrm{M} \rightarrow 1$ and therefore all definitions are similar.
    Today at $z=0$, there is a difference of a factor of around 3 between the 200m and the 180c definitions with the definitions of E98 (equation \ref{eq:E98}) and BN98 (equation \ref{eq:BN98}) in the middle.}
    \label{fig:rhoSO}
  \end{center}
\end{figure}

Often used definitions in practice are $\rho_\mathrm{SO} = 200 \, \rho_\mathrm{mean}$ (200m), $\rho_\mathrm{SO} = 180 \, \rho_\mathrm{crit}$ (180c) or one of the definitions already given in equation (\ref{eq:E98}) (E98) or (\ref{eq:BN98}) (BN98) above.
In Fig. \ref{fig:rhoSO}, we show the evolution of those 4 commonly used definitions over a range of 10 Gyr in a $\Lambda$CDM universe with $\Omega_\mathrm{M,0} = 0.28$ and $\Omega_{\Lambda,0} = 0.72$.
At early times (high redshift), the contribution of a cosmological constant is vanishing, $\Omega_\mathrm{M} \rightarrow 1$ and as a consequence $\rho_\mathrm{mean} \rightarrow \rho_\mathrm{crit}$.
Therefore, all of those 4 spherical overdensity definitions are similar at high redshift.
Today at $z=0$, there is a difference of a factor of around 3 between the 200m and the 180c definitions with the E98 and BN98 definitions in the middle.\footnote{Both fits to the generalized spherical collapse model agree quite well in this time range but the form of the fitting function in equation (\ref{eq:E98}) is much simpler.}
The overall evolution of all 4 definitions over approximately the last 10 Gyr is quite evident from Fig. \ref{fig:rhoSO}.
In the literature, any of these 4 definitions, and other similar definitions, are used as proxy to define a virialized region.
For example in the case 200m, the formal virial radius and mass are defined by $r_\mathrm{vir} = r_\mathrm{200m}$ and $M_\mathrm{vir} = M_\mathrm{200m} = M(r_\mathrm{200m})$, respectively.
In appendix \ref{sec:massratio}, we illustrate the evolution of the difference between the various formal virial mass definitions used in this study.

Here, it is interesting to remark that not all of those definitions will be physically meaningful in the far future.
In the future, the critical density will approach the asymptotic value $\rho_\mathrm{crit,\infty} = 3 H_0^2 \Omega_{\Lambda,0}/8 \pi G \, (= 97.91 \, \Mo \, \mathrm{kpc}^{-3}$ in our cosmology), whereas the mean matter density $\rho_\mathrm{mean}$ goes to zero.
Therefore, the 180c and BN98 definitions will reach asymptotic values and are well defined also in the far future.
The 200m and E98 spherical overdensity thresholds approach zero and will enclose larger and larger regions.
The enclosed mass in these cases will converge but is typically larger than the ultimate bound mass within the turnaround radius \cite[see e.g.][]{2005MNRAS.363L..11B}.
Of course, the E98 and BN98 fits to the generalized spherical collapse model are only valid for a certain range of scale factors (for details consult the papers) and should not be extrapolated into the far future.

\section{Evolution of physical properties}\label{sec:property_evolution}

\subsection{Simulation data}

\begin{deluxetable*}{lccc}
  \tablecaption{Number of all (left) and complete (right) halo tracks over the last 10 Gyr in each mass class for the 3 different size/mass definitions.
  The halo tracks are assigned to mass classes according to the mass $M_0$ the halo has at $z=0$ for each size/mass definition, i.e. $M_\mathrm{200m,0}$, $M_\mathrm{E98,0}$ respectively $M_\mathrm{180c,0}$.\label{tab:htn}}  
  \tablehead{\colhead{Mass class} & \colhead{200m} & \colhead{E98} & \colhead{180c}}
  \startdata
    $10.0 \leq \log_{10}(M_0/\Mo) < 10.5$ & 2988 / 2454 & 2877 / 2434 & 2795 / 2424 \\
    $10.5 \leq \log_{10}(M_0/\Mo) < 11.0$ & 1241 / 1168 & 1198 / 1140 & 1127 / 1089 \\
    $11.0 \leq \log_{10}(M_0/\Mo) < 11.5$ &  456 /  453 &  404 /  399 &  366 /  362 \\
    $11.5 \leq \log_{10}(M_0/\Mo) < 12.0$ &  175 /  173 &  158 /  157 &  150 /  149 \\
    $12.0 \leq \log_{10}(M_0/\Mo) < 12.5$ &   57 /   57 &   55 /   55 &   48 /   48 \\
    $12.5 \leq \log_{10}(M_0/\Mo) < 13.0$ &   29 /   29 &   29 /   29 &   26 /   26 \\
    $13.0 \leq \log_{10}(M_0/\Mo) < 13.5$ &   10 /    9 &    6 /    5 &    6 /    5 \\
    $13.5 \leq \log_{10}(M_0/\Mo)$        &    3 /    3 &    3 /    3 &    3 /    3 \\
  \enddata
\end{deluxetable*}

In order to illustrate some points about the evolution of physical properties, we use data from a moderately sized dissipationless dark matter N-body simulation that was used in order to select objects for the hydrodynamical simulations presented in \cite{2012ApJ...748...54Z}.
The periodic simulation box had a comoving length of $L_\mathrm{box} = 25.6 \, h^{-1} \, \mathrm{Mpc} \approx 36.57 \, \mathrm{Mpc}$ and the adopted $\Lambda$CDM cosmology had a total matter density parameter $\Omega_\mathrm{M,0} = 0.28$, cosmological constant $\Omega_{\Lambda,0} = 0.72$, Hubble parameter $H_0 = 100 \, h \, \mathrm{km} \, \mathrm{s}^{-1} \, \mathrm{Mpc}^{-1}$, with $h = 0.7$, linearly extrapolated normalization of the power spectrum $\sigma_8 = 0.82$, and spectral index $n_s = 0.96$.
The number of particles was $256^3 \approx 1.678 \times 10^{7}$ resulting in a particle mass of $1.110 \times 10^{8} \, \Mo$ and the softening for each particle was set to 1/20 of the initial inter-particle separation resulting in a softening length of $\epsilon$ = 7.143 kpc.
The time evolution was performed with the N-body code PKDGRAV2 \citep{2001PhDT........21S}.

At 8 snapshots equally spaced in cosmic time over the last 10 Gyr before $z=0$ (the exact time range is 9.609 Gyr, $z=1.592 \ldots 0$), we have identified the dark matter halos with the AMIGA Halo Finder (AHF, v1.0-064)\footnote{\texttt{http://popia.ft.uam.es/AHF/}} \citep{2009ApJS..182..608K} and tracked the individual halos over the different epochs with the merger tree tool that is part of the AHF package.
The resulting halo catalogs and tracks allow us to study the properties of individual dark matter halos and their evolution with time.

For each halo at the various epochs, we generate spherically averaged profiles from 2.5 $\epsilon$ out to several Mpc with the profiling tool presented in the appendix of \cite{2012ApJ...748...54Z}.
The bins are logarithmically spaced and we use 10 bins per dex.
We run the analysis for 3 different spherical overdensity size/mass definitions: 1) 200m, 2) E98 and 3) 180c.
In each case, we only use halos which are distinct, i.e. over the whole time the halo was never part of a larger halo or got tidally truncated.
We allow for only partially complete halo tracks over the considered time range, i.e. we include halos that form at a redshift lower than $z=1.592$.
However, for most of the halos the track is complete over the considered time range.

We then classify the halo tracks according to the mass the halo has at $z=0$ in each of the 3 size/mass definitions and calculate the median as well as the 15th and 85th percentile over all halo tracks in each bin.
With this we can make statistically more meaningful statements about the typical behavior of a class of halos.
We have checked and using only complete halo tracks makes hardly any difference in the medians and percentiles.
In Table \ref{tab:htn} we summarize the number of halo tracks in each mass bin.

The most massive object in our simulation has a mass at $z=0$ of ($M_\mathrm{200m,0},M_\mathrm{E98,0},M_\mathrm{180c,0}) = (2.957,2.201,1.652) \times 10^{14} \, \Mo$.
Thus, we only have a few objects with masses above $10^{13.5} \, \Mo$ for all 3 mass definitions within our simulation box.
Therefore, we can only indicate trends for those clusters which are still in the assembly process today.
Our focus here is mainly on objects smaller than clusters anyway.
Those objects have typically a rather quiet recent merger history and do not show much evolution as we will see later.

We present the various evolution plots as a function of cosmic time $t$.
The linearity of time is more intuitive for human beings from daily experience - at least we feel this way.
Time expressed as redshift $z$ or scale factor $a$ can often give misleading impressions about how fast physical processes run in cosmology.
Nonetheless, we plot on the top axis the corresponding redshift as well for those who want to compare with the more observer-friendly redshift as time indicator.

\subsection{Spherical overdensity mass and size evolution}

\begin{figure*}
  \begin{center}
    \includegraphics[width=\textwidth]{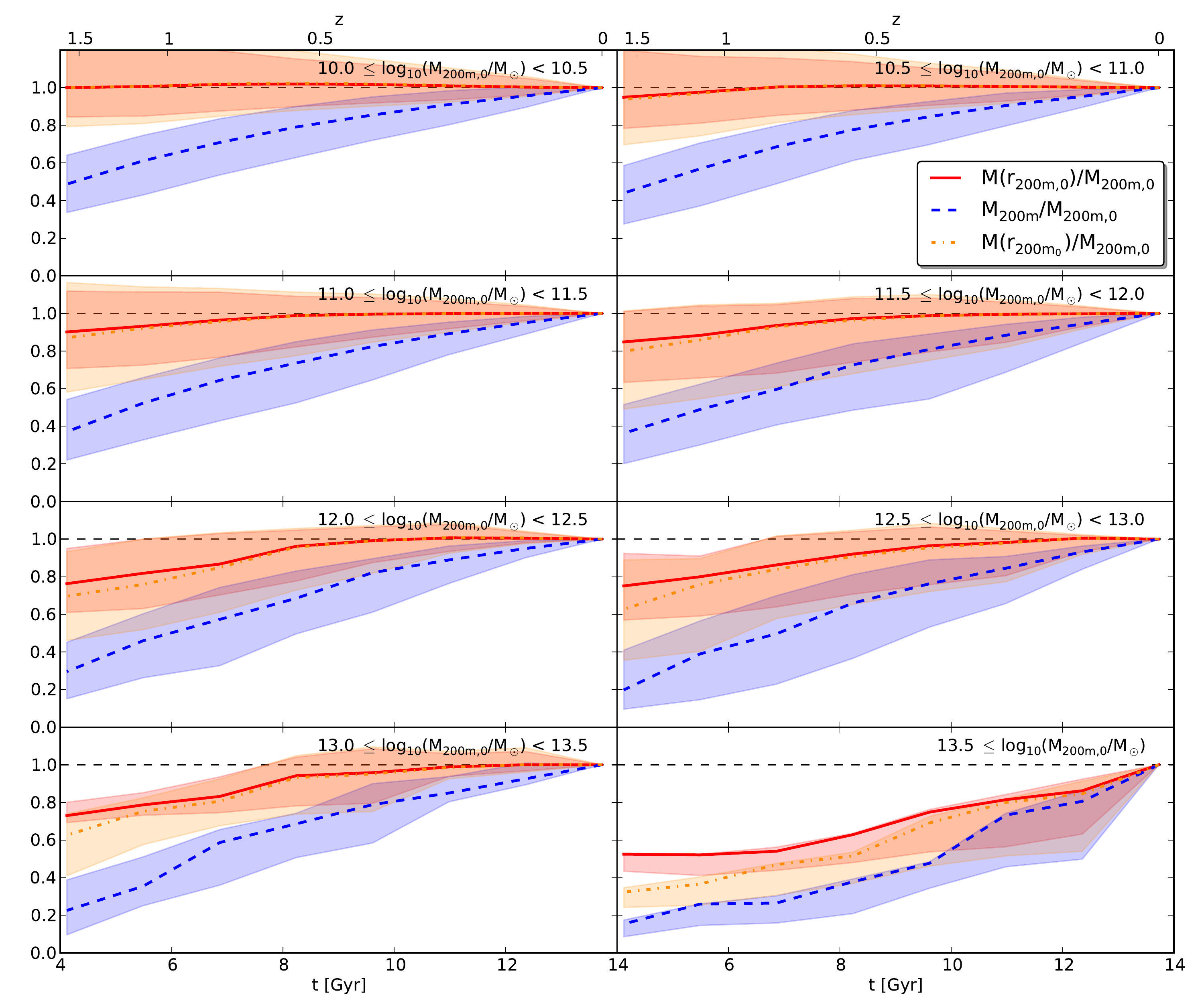}
    \caption{Mass evolution for the size/mass definition 200m, $M_\mathrm{200m}$, over the last 10 Gyr for different classes of halos.
    Also shown is the evolution of the mass within a fixed physical radius of $r_\mathrm{200m,0}$, the size at $z=0$, as well as the alternative size/mass definition where the spherical overdensity $\rho_\mathrm{SO}$ was fixed at the value at $z=0$ in order to find the size $r_\mathrm{200m_0}$.
    For each halo track the masses are normalized by the mass of the halo at $z=0$, $M_\mathrm{200m,0}$.
    Plotted is the median (lines) over all halo tracks in the various mass bins as well as the 15th to 85th percentile range (shaded regions).
    It is clearly evident that the spherical overdensity size/mass definition overpredicts the growth when compared to the mass content within a fixed physical scale.}
    \label{fig:mass_evolution}
  \end{center}
\end{figure*}

In Fig. \ref{fig:mass_evolution}, we show the mass evolution for the size/mass definition 200m, $M_\mathrm{200m}$, over the last ca. 10 Gyr for different classes of halos as specified in Table \ref{tab:htn}.
For comparison, we also plot the evolution of the mass within a fixed physical scale of $r_\mathrm{200m,0}$, the size each halo had at $z=0$.
For each halo track the masses are normalized by the mass of the halo at $z=0$, $M_\mathrm{200m,0}$.
It can be easily seen that the spherical overdensity size/mass definition overpredicts the growth when compared to the mass content within a fixed physical scale. For example, for halos that are in the mass range $10.0 \leq \log_{10}(M_\mathrm{200m,0}/\Mo) < 10.5$ one would assume that they have grown by a factor of 2 in the last 10 Gyr when looking at $M_\mathrm{200m}$ but typically all of the mass was essentially already in place 10 Gyr ago.
Similarly for the mass range $12.0 \leq \log_{10}(M_\mathrm{200m,0}/\Mo) < 12.5$ where one would traditionally come to the conclusion that those objects have typically grown by ca. a factor of 3 whereas $\approx 75\%$ of the mass was already in place within $r_\mathrm{200m,0}$ 10 Gyr ago.

This discrepancy is clearly due to the evolving spherical overdensity threshold $\rho_\mathrm{SO}$ as shown in Fig. \ref{fig:rhoSO}.
If we fix this threshold density to the value it has at $z=0$ and then calculate the mass within the resulting sizes $r_\mathrm{200m_0}$, we can see in Fig. \ref{fig:mass_evolution} that this definition is essentially identical to the evolution of the mass within the fixed physical scale $r_\mathrm{200m,0}$.
Only for halos in higher-mass bins do we see some deviation mainly at high redshift.

\begin{figure}
  \begin{center}
    \includegraphics[width=\columnwidth]{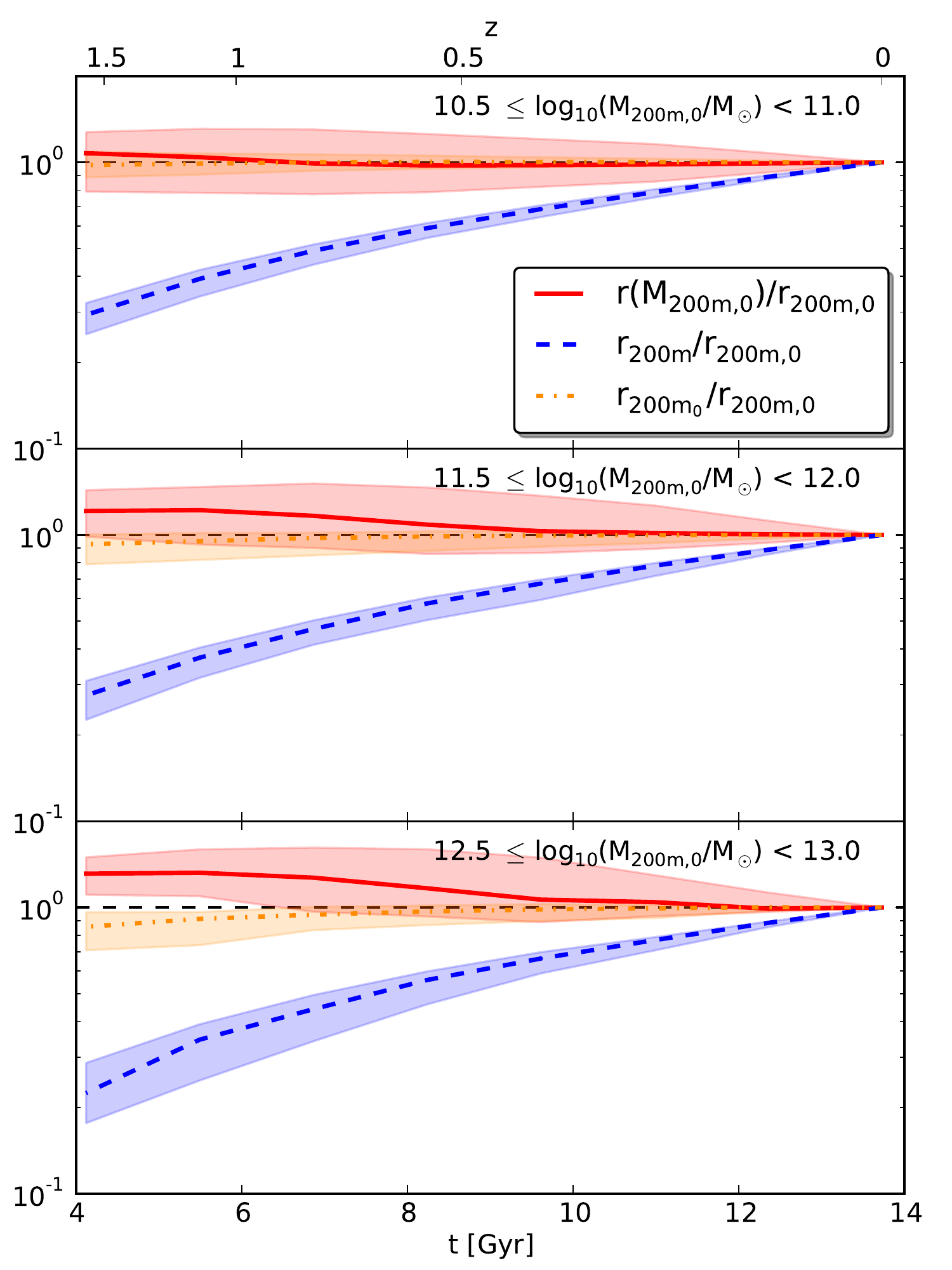}
    \caption{Relative size evolution for different definitions over the last 10 Gyr for a subset of the mass categories.
    Plotted are medians over all halo tracks as lines as well as the range between the 15th and 85th percentiles as shaded region.
    Reflecting the large growth in mass seen before, we can also observe an extensive increase in the associated size $r_\mathrm{200m}$.
    We also see that the radius that contains the mass at $z=0$, $r(M_\mathrm{200m,0})$, is still contracting for higher-mass halos at high redshift.
    Later on, however, and for low-mass objects that mass shell is essentially in place and does not vary much with time.
    Similarly the scale $r_\mathrm{200m_0}$ resulting from keeping the spherical overdensity threshold constant shows only some growth early on for high-mass objects.}
    \label{fig:size_evolution}
  \end{center}
\end{figure}

In Fig. \ref{fig:size_evolution}, a complementary version of Fig. \ref{fig:mass_evolution} for sizes, we present for a subset of the mass categories (for brevity) the evolution of the different size scales associated with the different definitions.
The large growth in mass seen before is also reflected in the extensive increase in the associated size $r_\mathrm{200m}$.
We also see that the radius that contains the mass at $z=0$, $r(M_\mathrm{200m,0})$, is still contracting for higher-mass halos at high redshift.
Later on, however, and for low-mass objects that mass shell is essentially in place and does not vary much with time.
Similarly, the scale $r_\mathrm{200m_0}$ resulting from keeping the spherical overdensity threshold constant shows only some growth early on for high-mass objects.

This is a clear indication, that for all halo classes there was much more matter already in place 10 Gyr ago than one would infer from the usual spherical overdensity definition.
For small objects there seems to be not much physical accretion at all and the amount of physical accretion increases with halo mass.
However, for all halo mass bins the amount of growth is overpredicted when compared to what happens within fixed physical scales.
This pseudo-evolution due to the evolving density threshold leads to an illusion of growth for dark matter halos.

\begin{figure}
  \begin{center}
    \includegraphics[width=\columnwidth]{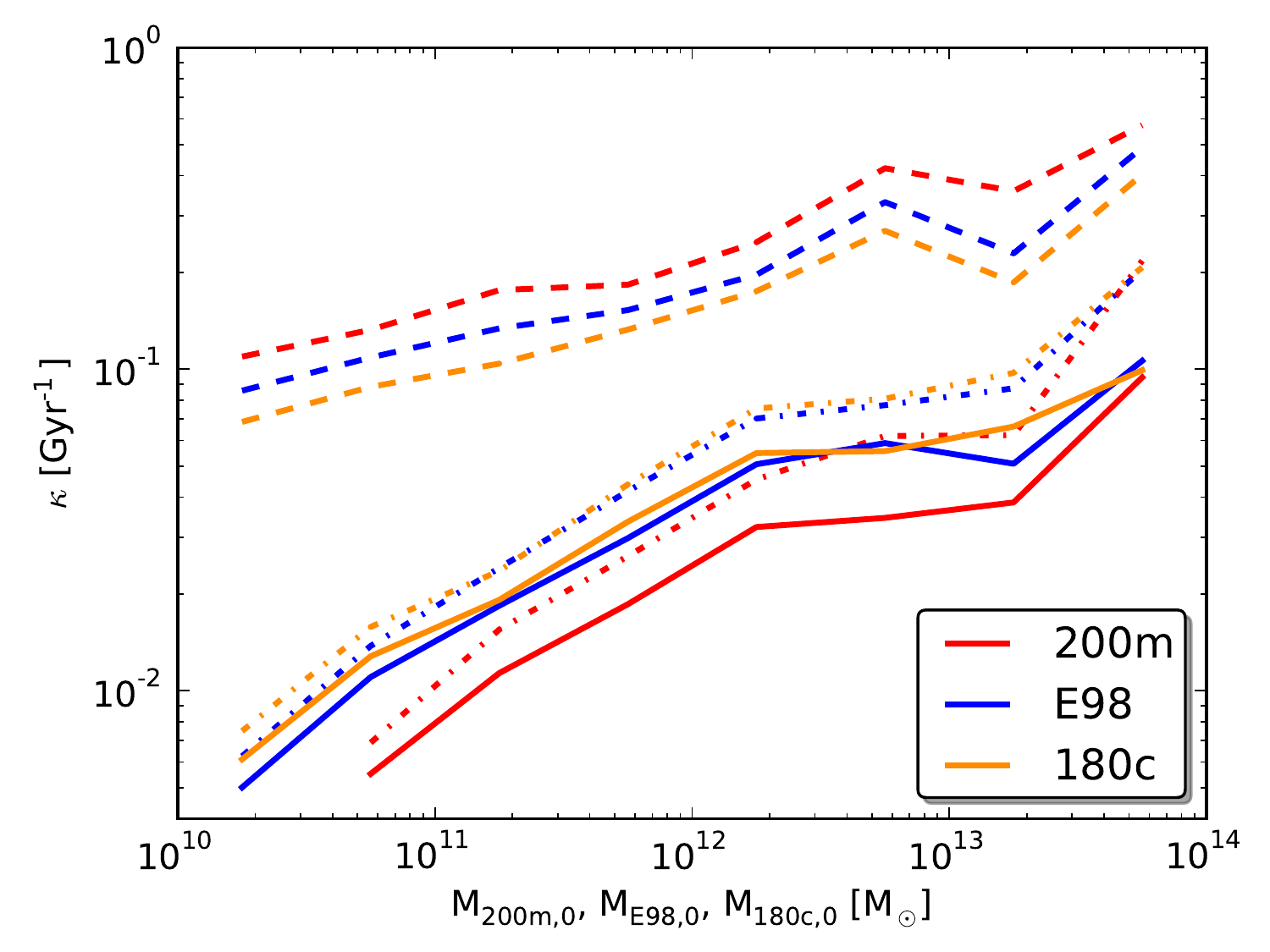}
    \caption{Average relative growth rate of the various mass medians as in Fig. \ref{fig:mass_evolution} for all 3 different size/mass definitions.
    As in Fig. \ref{fig:mass_evolution}, 3 types of medians are shown (same line style): radius fixed at $z=0$ value (solid, e.g. $M(r_\mathrm{200m,0})/M_\mathrm{200m,0}$), evolving density threshold (dashed, e.g. $M_\mathrm{200m}/M_\mathrm{200m,0}$), and density threshold fixed at $z=0$ value (dash-dotted, e.g. $M(r_\mathrm{200m_0})/M_\mathrm{200m,0}$).}
    \label{fig:mass_evolution_relative}
  \end{center}
\end{figure}

In order to present our findings in a compact way and to include the results from the other 2 size/mass definitions, we show the average relative growth rate of the various mass medians of Fig. \ref{fig:mass_evolution} for all 3 size/mass definitions as a function of halo class in Fig. \ref{fig:mass_evolution_relative}.
For every mass bin we calculate for each median curve $m$ in Fig. \ref{fig:mass_evolution}
\begin{equation}
  \kappa \equiv \frac{(m_0-m_{1.592})/m_{1.592}}{\Delta t}
\end{equation}
with $\Delta t = 9.609$ Gyr over the redshift range $z=1.592 \ldots 0$ in our cosmology.
We show the same 3 types of medians as already plotted in Fig. \ref{fig:mass_evolution} (same line style): radius fixed at $z=0$ value (solid, e.g. $M(r_\mathrm{200m,0})/M_\mathrm{200m,0}$), evolving density threshold (dashed, e.g. $M_\mathrm{200m}/M_\mathrm{200m,0}$), and density threshold fixed at $z=0$ value (dash-dotted, e.g. $M(r_\mathrm{200m_0})/M_\mathrm{200m,0}$).
Of course the growth is, in general, not linear with cosmic time over this time interval (as can be seen in Fig. \ref{fig:mass_evolution}) but this average linear growth rate serves as a good indicator to quantify the differences between the various definitions.

In general, the average relative growth rate $\kappa$ increases with halo mass for all mass definitions.
There is, however, quite a discrepancy between the evolving spherical overdensity definition (dashed) and the other 2 definitions where either the physical size or the spherical overdensity threshold was held constant.
The difference is larger the more the spherical overdensity threshold (see Fig. \ref{fig:rhoSO}) varies with time, i.e. largest for the 200m and smallest for the 180c definition.
Also, for small-mass halos the degree of pseudo-evolution is larger than for high-mass objects.\footnote{The attentive reader has probably realized, that for the 200m definition the data points for the 2 definitions, where the physical size or the spherical overdensity threshold was held constant, are missing in the smallest halo mass bin in Fig. \ref{fig:mass_evolution_relative}. This is due to the fact that for both cases the median growth is marginally negative in our data for this mass bin and hence an artifact of the plot having a logarithmic $y$-axis.}

\subsection{Intrinsic physical scales}

In this section, we would like to quantify the evolution of several intrinsic physical scales of dark matter halos.

\subsubsection{Maximum value of circular velocity curve}

\begin{figure}
  \begin{center}
    \includegraphics[width=\columnwidth]{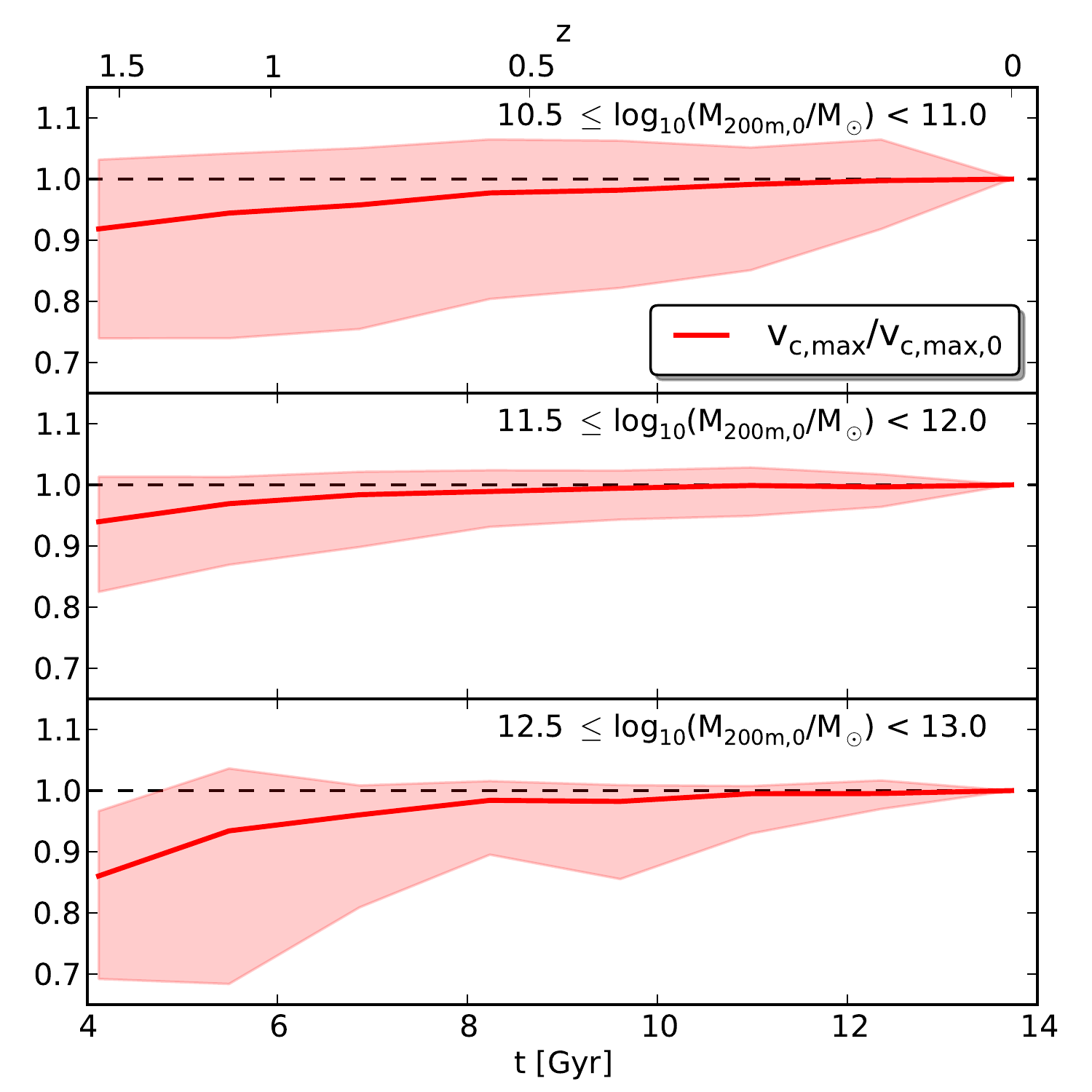}
    \caption{Relative evolution of the maximum value of the circular velocity curve, $v_\mathrm{c,max}$, over the last 10 Gyr.
    The same mass classes as in Fig. \ref{fig:size_evolution} are shown.
    The medians are plotted as lines and the range between the 15th and 85th percentiles is marked by the shaded region.
    The evolution is rather moderate over this time interval and reflects more the behavior already seen for the size and mass of halos when fixed physical scales are considered.}
    \label{fig:vcmax_evolution}
  \end{center}
\end{figure}

The maximum value of the circular velocity curve is an intrinsic physical scale for an individual dark matter halo.
The value of $v_\mathrm{c,max}$ is often used in the literature as an alternative to mass in order to characterize the 'size' of a dark matter halo.
In Fig. \ref{fig:vcmax_evolution}, we show the evolution of $v_\mathrm{c,max}$ over the last 10 Gyr.
The same mass bins as in Fig. \ref{fig:size_evolution} are shown.
We see that the overall evolution is rather moderate and much more reflects the modest evolution of size and mass of halos when fixed physical scales are considered (see Figs. \ref{fig:mass_evolution} and \ref{fig:size_evolution}).

As expected for a characteristic scale of a dark matter halo, the determination of the maximum value of the circular velocity curve is not sensitive to the size/mass definition used.
Therefore, the evolution of $v_\mathrm{c,max}$ differs only marginally for the other 2 size/mass definitions from what is shown in Fig. \ref{fig:vcmax_evolution} for the 200m definition.

\subsubsection{Scale radius}

\begin{figure}
  \begin{center}
    \includegraphics[width=\columnwidth]{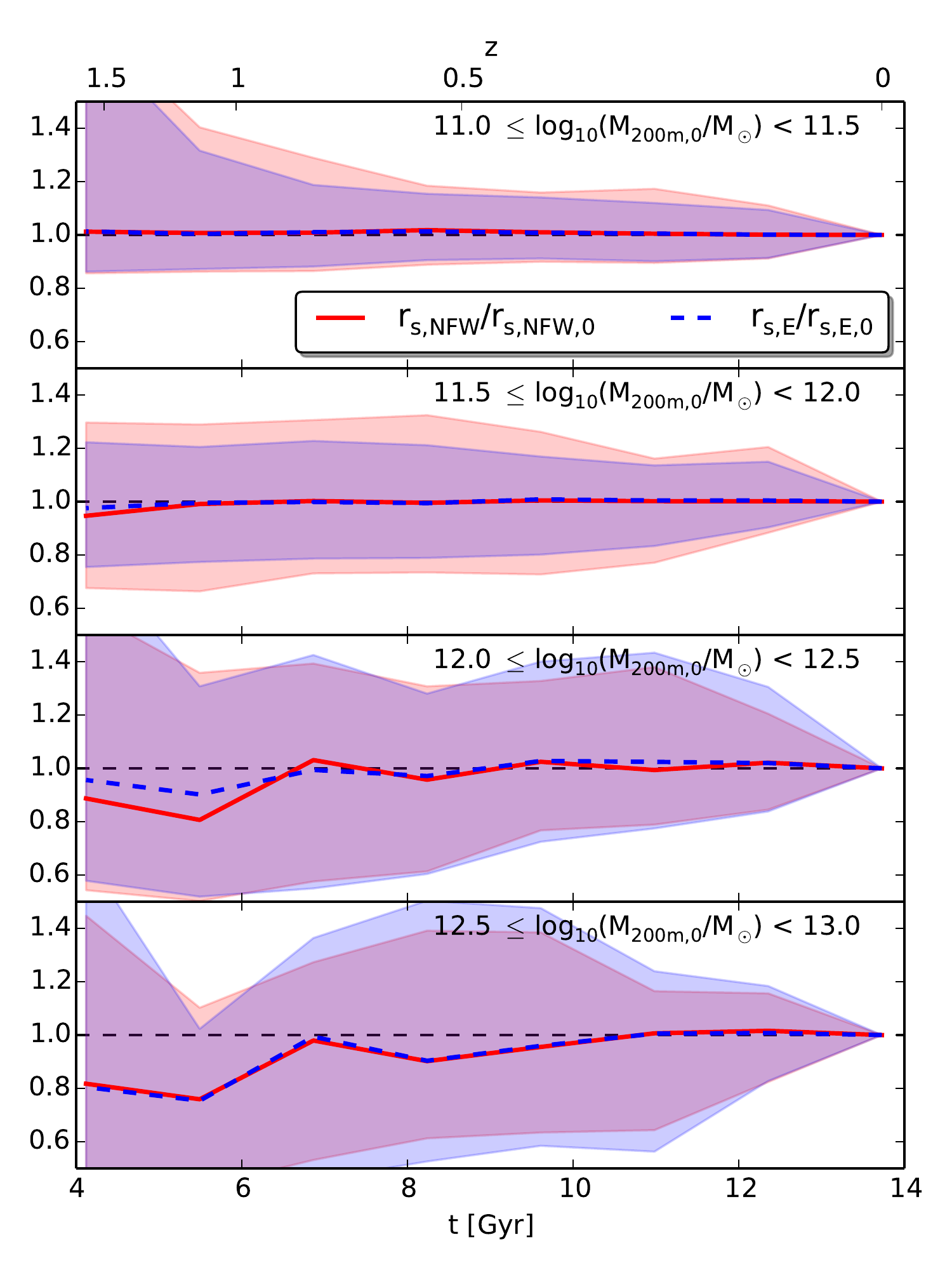}
    \caption{Evolution of the scale radius over the last 10 Gyr for the NFW as well as the Einasto fit for the 200m size/mass definition.
    The evolution is normalized to the values at $z=0$.
    The lines denote the medians and the shaded region the range between the 15th and 85th percentile.
    It is evident that for low-mass objects the scale radius $r_\mathrm{s}$ does not show much evolution.
    Even for the $12.5 \leq \log_{10}(M_\mathrm{200m,0}/\Mo) < 13.0$ mass class the median growth of $r_\mathrm{s}$ is rather moderate in both cases.}
    \label{fig:rs_evolution}
  \end{center}
\end{figure}

The family of the $\alpha\beta\gamma$ models, given by
\begin{equation}\label{eq:NFW}
  \rho(r) = \frac{\rho_0}{(r/r_\mathrm{s})^\gamma [1+(r/r_\mathrm{s})^\alpha]^\frac{\beta-\gamma}{\alpha}} \, ,
\end{equation}
is often used to describe the spherically averaged density structure of various objects in the universe.
\cite{1996ApJ...462..563N} found that the parameter set $(\alpha,\beta,\gamma) = (1,3,1)$ provides excellent fits for the density profiles of dark matter halos in dissipationless structure formation simulations: the so-called NFW profile.
Recent work has shown, however, that the central density profile is shallower than $\rho \propto r^{-1}$ \citep{2008MNRAS.391.1685S,2009MNRAS.398L..21S,2010MNRAS.402...21N} and that often the \cite{1965TrAlm...5...87E} profile
\begin{equation}\label{eq:E}
  \rho(r) = \rho_0 \exp\left(-\frac{2}{\alpha}\left[(r/r_\mathrm{s})^\alpha-1 \right]\right)
\end{equation}
provides a better fit.

A generic feature of both of these density profiles is that at a certain physical scale, the scale radius $r_\mathrm{s}$, a transition between a shallower inner region to a more steeper outer regions happens.
The dividing logarithmic slope at $r_\mathrm{s}$ is $\mathrm{d} \ln(\rho) / \mathrm{d} \ln(r) |_{r_\mathrm{s}} = -2$ for the NFW as well as the Einasto profile, i.e. $r_\mathrm{s} = r_{-2}$, the location where the logarithmic slope is $-2$.
Therefore, we would like to describe the evolution of this physical transition scale $r_\mathrm{s}$ with cosmic time.

For each dark matter halo we fit an NFW profile where the fitting radial range was limited to 0.8 times the size the halo had at $z=0$ (i.e. $r_\mathrm{200m,0}$, $r_\mathrm{E98,0}$ respectively $r_\mathrm{180c,0}$) at all times, i.e. the fitting range was constant in time.
It is common practice to limit the fitting range in order to avoid complications with massive substructures in the outer parts of dark matter halos \cite[see e.g.][]{2010MNRAS.402...21N}.
We have experimented with various other fitting ranges and our conclusions are not sensitive to the specific choice.
The Einasto profile has one more free parameter than the NFW profile.
We fixed the shape parameter to $\alpha=0.16$, which is similar to other recent work in the literature \cite[e.g.][]{2008MNRAS.391.1685S,2010MNRAS.402...21N}.
We have checked that using similar, reasonable values for $\alpha$ do not affect the fit values for $r_\mathrm{s}$ dramatically.
Therefore, our fitting procedure is quite robust and not sensitive to choices of parameters.\footnote{We even tried more methods to determine the transition radius $r_\mathrm{-2}$. A smoothing spline fit to the density profile or a $(\alpha,\beta,\gamma) = (1,2.5,1)$ profile fit give essentially the same results for the evolution of $r_\mathrm{-2}$. For clarity, we only present the results from the NFW and Einasto fit. Interestingly, the $(\alpha,\beta,\gamma) = (1,2.5,1)$ profile has the smallest scatter in the $r_{-2}$ evolution.}
The fitting was done with the non-linear least-square minimization Python package lmfit\footnote{\texttt{http://lmfit.github.io/lmfit-py/}} where we used the standard Levenberg-Marquardt algorithm in order to minimize the figure-of-merit or residual function
\begin{equation}
  R^{2} \equiv \sum_{i=1}^{N_\mathrm{bin}} \left( \ln(\rho_i^\mathrm{data}) - \ln(\rho_i^\mathrm{fit}\right)^{2} \, .
\end{equation}

In Fig. \ref{fig:rs_evolution}, we show the relative evolution of the scale radius for the NFW and Einasto profile fit for the 200m size/mass definition.
We only show mass bins where the formal virial radius $r_\mathrm{200m,0}$ contains at least ca. 1000 particles so that the internal structure of the halo is properly resolved.
For the low-mass halos there is nearly no evolution present whereas the $12.5 \leq \log_{10}(M_\mathrm{200m,0}/\Mo) < 13.0$ mass class shows some moderate growth over the last 10 Gyr.
The determination of the scale radius is not sensitive to the details of the size/mass definition, as expected for an intrinsic scale, and the results for the other 2 size/mass definitions are essentially the same as in Fig. \ref{fig:rs_evolution}.
Our results are in good agreement with previous results in the literature which also show that the scale radius is approximately constant in time for dark matter halos after their formation \cite[e.g.][]{2001MNRAS.321..559B,2003MNRAS.339...12Z}.

\begin{figure}
  \begin{center}
    \includegraphics[width=\columnwidth]{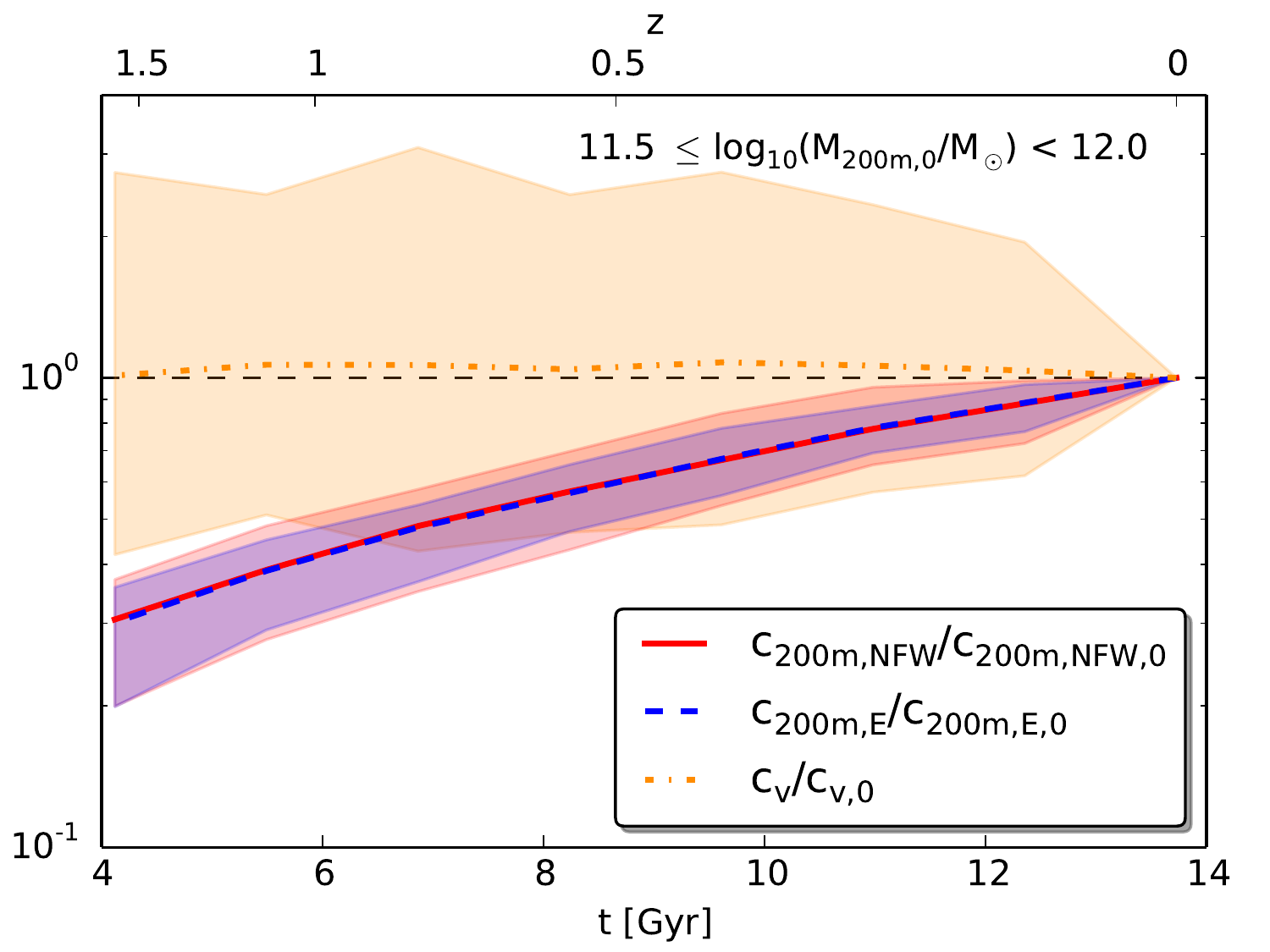}
    \caption{Relative evolution of the virial concentration for the NFW as well the Einasto profile fits with $r_\mathrm{vir} = r_\mathrm{200m}$.
    The range between the 15th and 85th percentile is marked by the shaded regions and the lines represent the medians.
    For Milky-Way-sized dark matter halos, one would think the virial concentration tripled in the last 10 Gyr.
    This evolution is completely driven by the artificial growth of the virial radius proxy $r_\mathrm{200m}$.
    Also shown is the alternative concentration measure $c_\mathrm{v}$, which is based on properties of the circular velocity curve (see equation \ref{eq:cv}) which essentially stays flat over the same time range, however with a rather large scatter (see main text for more details).}
    \label{fig:con_evolution}
  \end{center}
\end{figure}

An alternative way of describing the density profile of a dark matter halo is relating the scale radius $r_\mathrm{s}$ to the virial radius $r_\mathrm{vir}$ via the virial concentration parameter
\begin{equation}
  c_\mathrm{vir} \equiv r_\mathrm{vir}/r_\mathrm{s}
\end{equation}
where generally for the virial radius any spherical overdensity proxy is used (see Section \ref{sec:spherical_overdensity}).
In Fig. \ref{fig:con_evolution}, we show for brevity the evolution of the concentration parameter for roughly Milky-Way-sized objects ($11.5 \leq \log_{10}(M_\mathrm{200m,0}/\Mo) < 12.0$).
We show the median evolution for both the NFW as well as the Einasto profile fit.
Since the scale radius essentially stays constant (see Fig. \ref{fig:rs_evolution}), it is clear that nearly all the evolution is driven by the artificial growth of the virial radius proxy $r_\mathrm{200m}$ for this halo class and one would think that the concentration had tripled in the last 10 Gyr.
In \cite{2013ApJ...766...25D}, they find as well that especially for low-mass halos the concentration evolution is mainly driven by pseudo-evolution and we discuss this in more detail in Section \ref{sec:discussion}.

In Fig. \ref{fig:con_evolution}, we also present the evolution of an alternative concentration measure \citep{2002ApJ...572...34A,2007ApJ...667..859D}
\begin{equation}\label{eq:cv}
  c_\mathrm{v} \equiv \frac{\bar{\rho}(r_{v_\mathrm{c,max}})}{\rho_\mathrm{crit,0}} = 2 \left( \frac{v_\mathrm{c,max}}{H_0 \, r_{v_\mathrm{c,max}}}\right)^{2} \, .
\end{equation}
This intrinsic concentration measure has the advantage that it is well defined both for isolated halos and subhalos.
It does not make any assumptions about a specific shape of the density profile and does not depend on any arbitrarily defined outer edge of the halo.
This concentration measure stays rather flat over the same time range though with quite a large scatter.
Since the circular velocity curve is typically rather flat over some range, the location where the maximum circular velocity is reached, $r_{v_\mathrm{c,max}}$, can exhibit quite some scatter over time which is then reflected in the scatter of $c_\mathrm{v}$.
The large scatter is certainly a drawback for $c_\mathrm{v}$ but the median behavior nonetheless indicates that a concentration measure based on intrinsic properties of the circular velocity curve of the dark matter halo, $c_\mathrm{v}$, shows qualitatively a different behavior than the virial concentration $c_\mathrm{vir}$.

\subsubsection{Virialized region}\label{sec:virialscale}

\begin{figure}
  \begin{center}
    \includegraphics[width=\columnwidth]{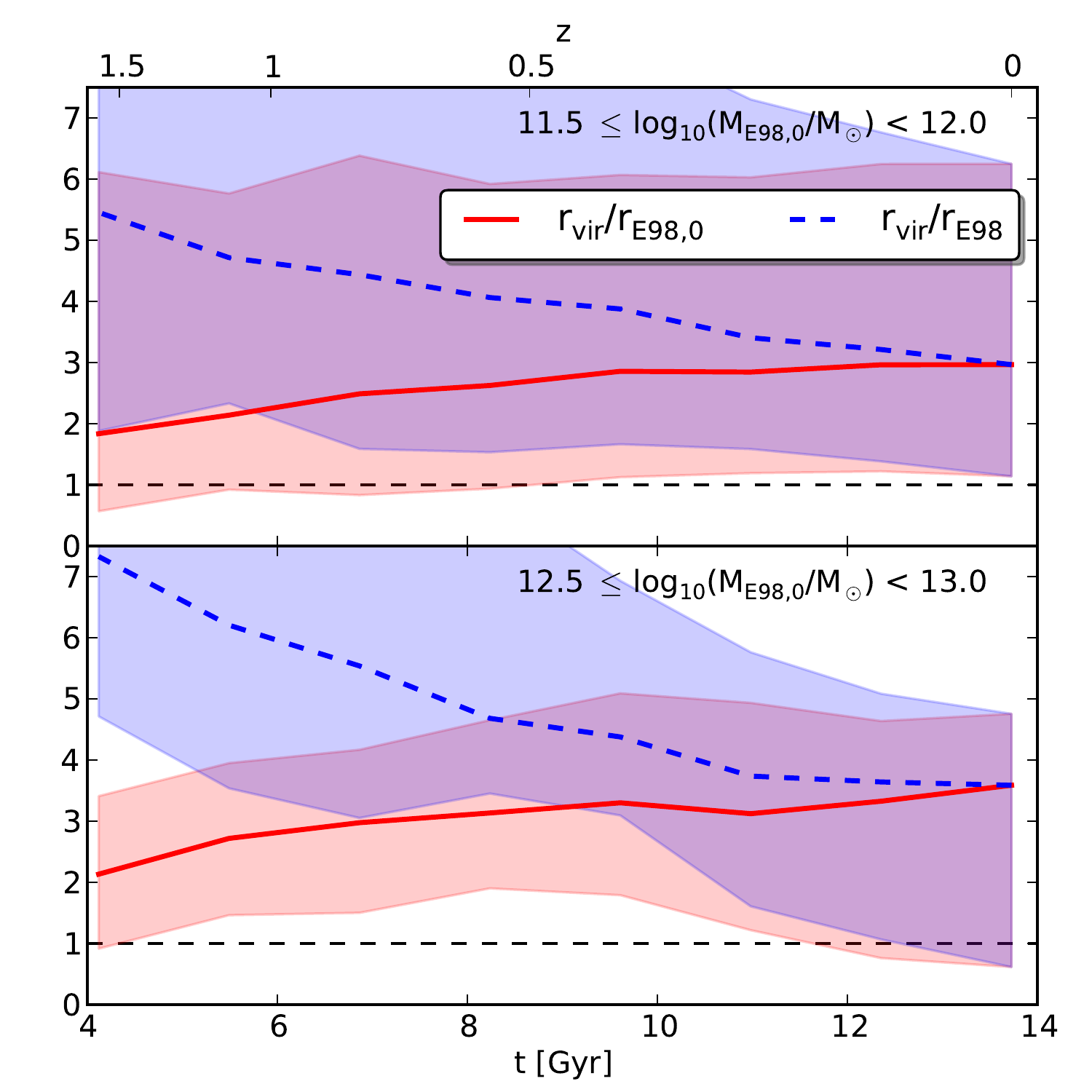}
    \caption{Evolution of our virial radius estimate, operationally defined via the virial ratio (see main text for details).
    In general, the virialized region is growing with time.
    The E98 definition is usually underestimating the virialized region.}
    \label{fig:virial_size_evolution}
  \end{center}
\end{figure}

In this section, we would like to explore the possibility to estimate the virialized region via an operational definition.
A natural choice is to look at the local virial ratio
\begin{equation}
  Q \equiv -\frac{2 K}{W}
\end{equation}
in each spherical shell of our profiles, where $K$ is the kinetic and $W$ potential energy of the matter in the shell (see Appendix \ref{sec:virialdetails} for more information on how we calculate $Q$ in detail).
From the scalar viral theorem $2K+ W = 0$ follows that for virial equilibrium we get $Q = 1$ and the shell is unbound for $Q > 2$.
Here, we define the virialized region where $Q < Q_\mathrm{crit} = 1.4$.
Of course, this operational definition depends on the choice of the $Q_\mathrm{crit}$ value.
Similar values as here with $Q_\mathrm{crit} = 1.4$ are used in the literature \citep[e.g.][]{2007MNRAS.376..215B,2007MNRAS.381.1450N}.
We have checked that our qualitative conclusions do not change with similar choices for $Q_\mathrm{crit}$.

In Fig. \ref{fig:virial_size_evolution}, we show the evolution of this operationally defined virial radius $r_\mathrm{vir}$ with time.
We compare with the definition E98, which should be the appropriate collapse model for a $\Lambda$CDM cosmology and show the results for 2 different halo masses.
It is evident that the virialized region in general physically grows with time, i.e. the median $r_\mathrm{vir}/r_\mathrm{E98,0}$ curve increases.
However, the E98 definition is in general underestimating the virialized region.

\begin{figure}
  \begin{center}
    \includegraphics[width=\columnwidth]{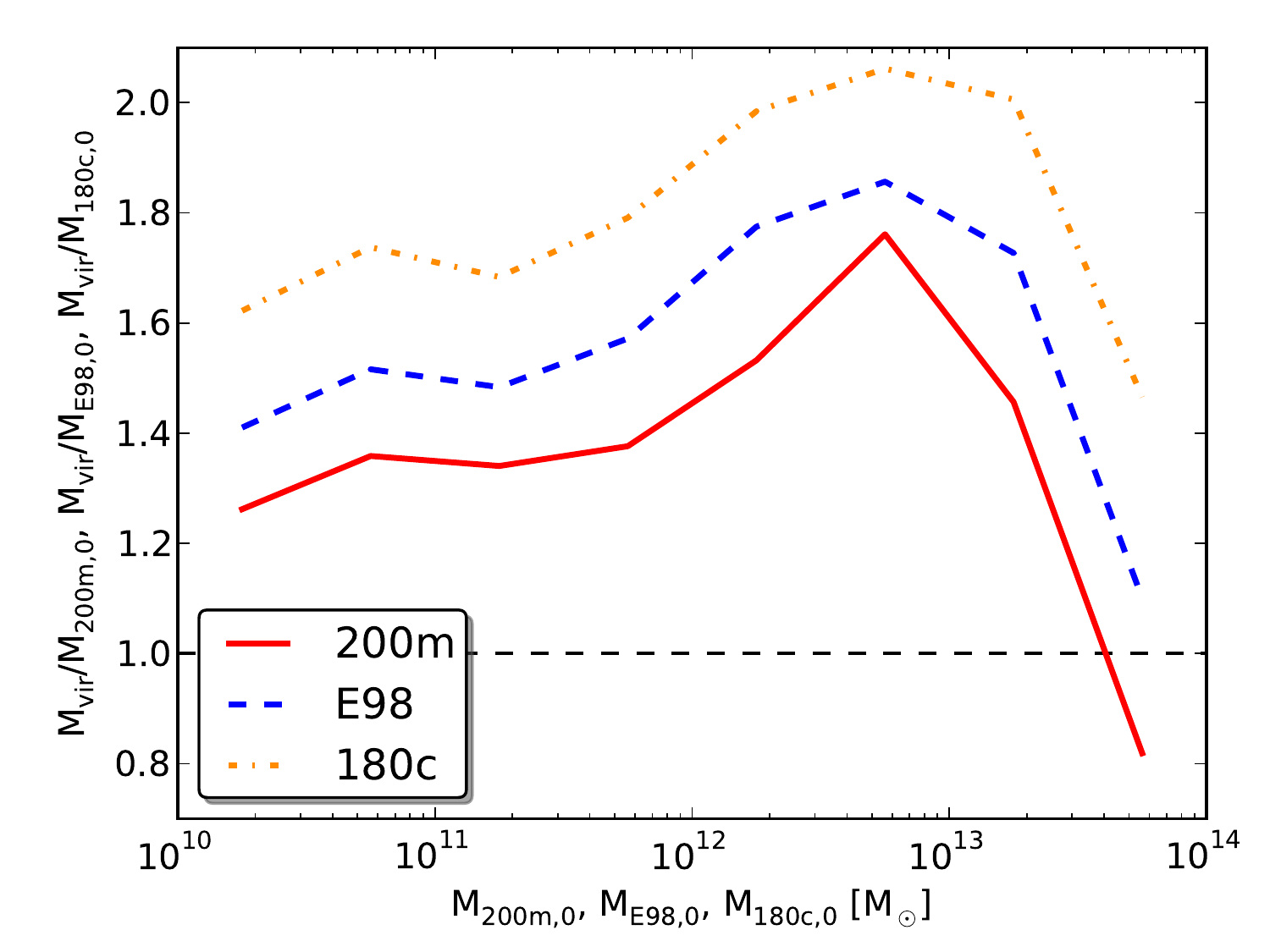}
    \caption{Median ratio of our operationally defined virial mass $M_\mathrm{vir}$ with the mass resulting from the 3 spherical overdensity definitions at $z=0$ as a function of halo mass.}
    \label{fig:virial_mass}
  \end{center}
\end{figure}

In Fig. \ref{fig:virial_mass}, the median ratio of the resulting virial mass $M_\mathrm{vir}$ with the mass from our 3 size/mass definitions are shown as a function of halo mass at $z=0$.
We find that for all masses and size/mass definitions, the operationally defined virial mass does not coincide with the spherical overdensity definition.
The characteristic shape of this ratio with the initial increase with halo mass and then decrease for higher-mass halos is due to the typical radial outflow (low-mass halos) respectively infall (high-mass halos) patterns around those dark matter halos.
For a more detailed discussion see \cite{2008MNRAS.389..385C} (figures 4 and 6) or the brief recapitulation below.

\cite{2003ApJ...588...35M} also used the differential virial ratio in order to define a boundary layer of a halo.
They, however, define the potential energy differently than we do in this study (see Appendix \ref{sec:virialdetails}).
They also find that virial masses determined with their method are larger than the E98 size/mass definition.
\cite{2008MNRAS.389..385C} used a different operational definition of the virialized region \citep[see also][]{2005MNRAS.363L..11B,2006ApJ...645.1001P}.
They use the mean radial velocity as a function of radius for a class of objects.
Typically, in central regions of dark matter halos that are in equilibrium, the radial velocity is close to zero.
Further out, the radial velocity curve starts deviating from zero.
For objects that are still accreting, the radial velocity is negative over some radial range further out due to the infalling matter.
Even further away, cosmic expansion manifests itself and the radial velocity becomes positive and grows with distance.
They found that small-mass objects do not have an infall region any more at $z=0$ and the radial velocity shows an outflow pattern beyond the central static region where the radial velocity is close to zero.
At earlier times, there was a moderate inflow region present in between the inner equilibrium and the outer cosmic expansion region for those small-mass objects which essentially has disappeared by $z=1$.
For more massive objects, such an infall region is still present today and it becomes more prominent with halo mass.
In the far future though, even for massive clusters, this infall pattern will disappear \citep{2005MNRAS.363L..11B}.
It is recommended to consult figures 4 and 13 in \cite{2008MNRAS.389..385C} which depict these patterns excellently.
They then define the virialized region as the central radial range where the radial velocity stays within a certain threshold from zero.
They call the size/mass associated with their definition the static radius $r_\mathrm{static}$ and static mass $M_\mathrm{static}$, respectively.
Their findings for the static size/mass are in good qualitative agreement with our results.\footnote{We also looked at the static mass and can confirm their findings. Thus, we opted to present a different approach to the operational definition of the virialized region based on the local virial ratio in this study.}
For example, they find a similar behavior of their static mass as we do in Fig. \ref{fig:virial_mass} as well as a general discrepancy between $M_\mathrm{static}$ and $M_\mathrm{BN98}$, which they use as their standard spherical overdensity definition (see their figures 6, 9, 14 and 15).

The aim here is certainly not to establish or promote a better way to define the virialized region of a dark matter halo - actually rather the contrary as we will discuss in Section \ref{sec:discussion}.
All operationally defined estimates depend on some choice of parameters as well assumed simplifications and symmetries (e.g. spherical symmetry).
Additionally, individual dark matter property profiles can be quite noisy if one is interested in local (differential) properties.
This can lead to quite some scatter (as in our case) and one can in general only make meaningful statements in a statistical sense by averaging over a class of halos than for an individual halo.
This is certainly a drawback of such methods when compared to the simple spherical overdensity size/mass definition that operates with a smoother, integrated quantity, namely the cumulative mass profile.
Also, such definitions can only be applied to simulation data where the full phase-space information is available.
Ultimately, we have to compare our results to observational data where such operational definitions of virialized regions are not feasible.

The main point we would make here is that, in general, a simple spherical overdensity criterion does not imply deeper fundamental physical properties like virialization.
A physically more precise nomenclature would be to name such size/mass definitions by their characteristic definition (e.g. $r_{200m}$ or $M_\mathrm{200m}$) and refrain from the practice of calling them virial radius or virial mass, since this is a deeper inherent physical property that does not follow from simple overdensity based size/mass definitions.

\subsection{Halo build-up}\label{sec:halobuildup}

\begin{figure*}
  \begin{center}
    \includegraphics[width=\textwidth]{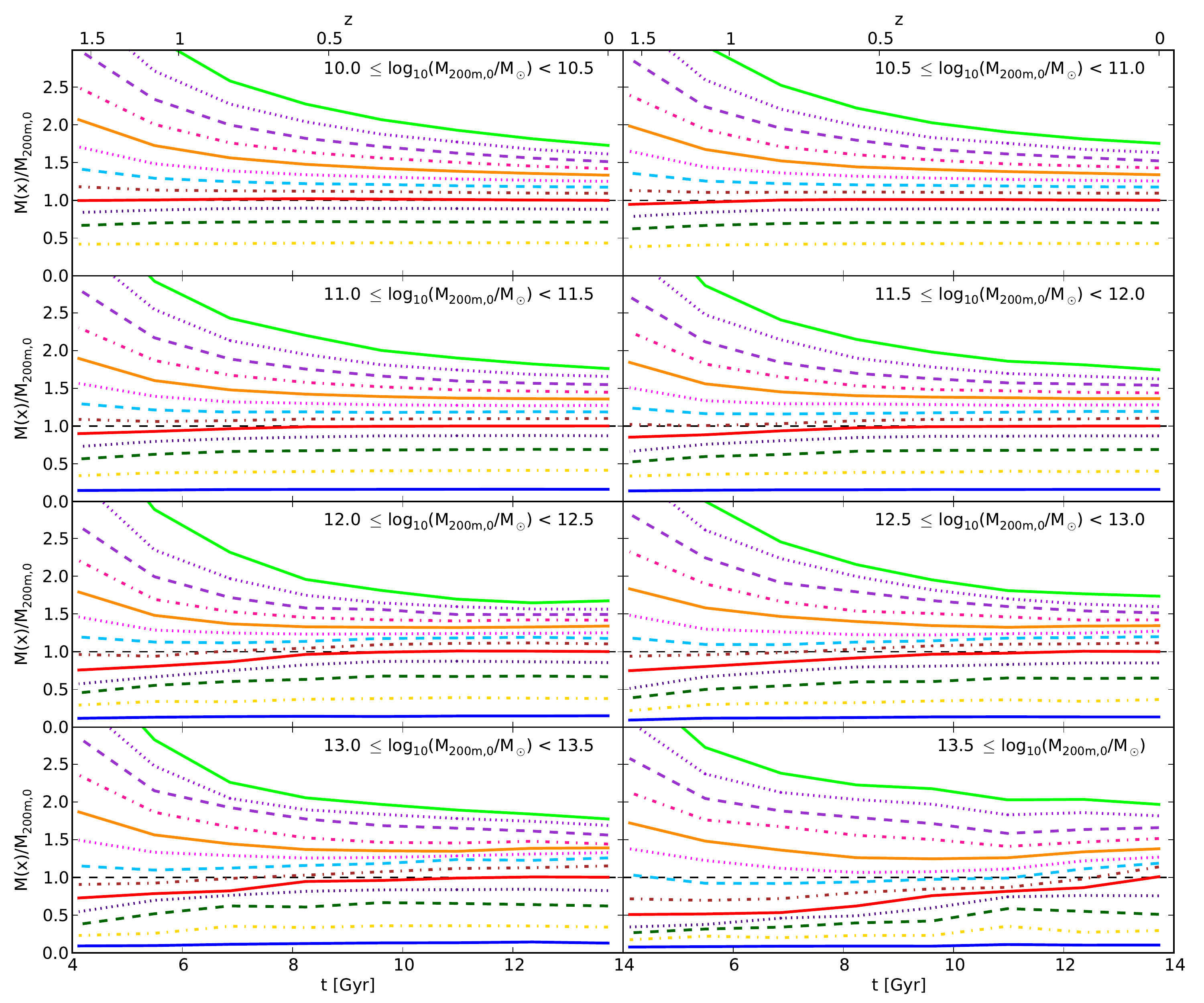}
    \caption{Halo build-up over the last 10 Gyr.
    Plotted are the medians of mass fractions $M(x)/M_\mathrm{200m,0}$ within various constant physical radii $x \equiv r/r_\mathrm{200m,0} = 3, 2.75, 2.5, 2.25, 2, 1.75, 1.5, 1.25, 1,0.75, 0.5, 0.25, 0.1$ (from top to bottom).
    For the smallest two mass classes, the $x=0.1$ line is missing since this is below the resolution scale in our simulation box.
    In general, the inner region of a halo was already in place 10 Gyr ago (see e.g. the $x=0.1$ and $x=0.25$ lines).
    For small-mass halos, the region out to $r_\mathrm{200m,0}$ also does not show much evolution whereas for more massive objects the growth of the outer regions is evident.
    Beyond $r_\mathrm{200m,0}$ the mass within a fixed physical radius is decreasing at early times.
    This is a manifestation of the Hubble flow.
    However, the mass shells beyond $r_\mathrm{200m,0}$ are not showing much evolution in the last few Gyr, except for the most massive objects.
    This illustrates nicely that the mass profile around dark matter halos extends way beyond a formal virial radius.}
    \label{fig:halo_buildup}
  \end{center}
\end{figure*}

In this section, we would like to illustrate how a halo of a certain mass class typically builds up over time.
In Fig. \ref{fig:halo_buildup}, we plot the medians over all halo tracks within a halo class of mass fractions $M(x)/M_\mathrm{200m,0}$ within various constant physical radii $x \equiv r/r_\mathrm{200m,0} = 3, 2.75, 2.5, 2.25, 2, 1.75, 1.5, 1.25, 1, 0.75, 0.5, 0.25, 0.1$ (from top to bottom).
The $x=0.1$ line is missing for the smallest 2 mass classes since this is below the resolution scale in our simulation box.

As evident from Fig. \ref{fig:halo_buildup}, the inner region of a halo was typically already in place 10 Gyr ago (see e.g. the $x=0.1$ and $x=0.25$ median lines).
The region out to $r_\mathrm{200m,0}$ does not show much evolution as well for small-mass halos.
For more massive objects, the mass assembly in the outer region is clearly visible.
At early times beyond $r_\mathrm{200m,0}$, the Hubble flow manifests itself since the mass within a fixed physical radius is decreasing.
In general, closer mass shells stabilize earlier than more distant mass shells.
However, during the last few Gyr the mass shells beyond $r_\mathrm{200m,0}$ do not show much evolution for all halo classes except for the most massive objects we consider.
Those objects are still in the assembly process today.
This is a good illustration that the mass profiles around dark matter density peaks extend way beyond a formal virial radius and do not show much evolution once a specific mass shell stabilized after the decoupling from the cosmic expansion (see also discussion in Section \ref{sec:discussion}).

This picture of the halo build-up is nicely complemented by the studies of \cite{2006ApJ...645.1001P}, \cite{2008MNRAS.389..385C} and \cite{2014ApJ...789....1D}, where they have studied the radial velocity patterns around dark matter halos (see also Section \ref{sec:virialscale}).
Low-mass objects today (i.e. low sigma peaks in the density field) stopped accreting a while ago since their infall region disappeared and the matter distribution about those peaks has essentially stabilized even beyond the formal virial radius.
More massive objects like, e.g., clusters (i.e. high sigma peaks) still have an infall region at $z=0$ and are therefore physically growing today.
Figures 4 and 13 in \cite{2008MNRAS.389..385C} or figure 8 in \cite{2014ApJ...789....1D} depict these radial velocity patterns excellently.

\section{Discussion}\label{sec:discussion}

In the previous Section \ref{sec:property_evolution}, we have seen that the evolution of spherical overdensity sizes and masses can lead to an inaccurate and distorted picture of the assembly process of dark matter halos.
Especially for low-mass objects, most of the physical matter distribution was already in place 10 Gyr ago (see also appendix \ref{sec:mechanism}).
Aspects of this pseudo-evolution effect have been pointed out before in the literature.
In \cite{2007ApJ...667..859D} they showed for a Milky-Way-sized object, the Via Lactea I halo, that after the last major merger at $z\approx 1.7$ the density profile is essentially established and the mass within fixed physical scales is roughly constant whereas the formal virial mass $M_\mathrm{200m}$ is still growing (see their figures 3 and 4).
They also point out that the small degree of physical accretion is typical for Milky-Way-sized halos by looking at a larger sample of halos in a cosmological volume.
\cite{2008MNRAS.389..385C} also show that the dark matter profile of a simulated Milky-Way-sized object typically does not evolve after $z=1$ (see their figures 2 and 16).
However, their main focus is on the discrepancy between an operationally defined virial mass, which they call static mass, that is defined by the size of the region of a halo with zero radial velocity, and the BN98 virial mass definition.
They found that the BN98 virial mass definition is, in general, underestimating/overestimating the static mass at low/high redshift (see their figures 15 and 16).
The transition mass scale where $M_\mathrm{static} = M_\mathrm{BN98}$ decreases dramatically with redshift so that especially for low-mass objects the associated dynamic mass of a halo is already underestimated for a long time.
This is similar to our findings in Section \ref{sec:virialscale} for an operational definition of the virial mass based on the local virial ratio.
This effect can be understood by the typical shape of the radial velocity profile for the different halo classes (see also Section \ref{sec:virialscale}).
Thus in general, the static mass does not coincide with the BN98 virial mass proxy.
\cite{2013ApJ...766...25D} looked at the pseudo-evolution effect in a statistical sense for a wide range of halo masses in the Bolshoi simulation \citep{2011ApJ...740..102K}.
They investigate the differences between the density profiles of dark matter halos between $z=1$ and $z=0$ and found that low-mass halos only grow by ca. 10\% on average during that time span.
By assuming a static halo density profile at $z=1/0$, they estimate as well the pseudo-evolution for each halo in 2 modes by forward/backward integration over this static profile with time.
We are tracking the individual halos for several simulation snapshots in cosmic time and quantify the pseudo-evolution effect differently by looking at fixed physical and intrinsic scales of the dark matter halos.
Their findings, however, are in good agreement with our results (see e.g. their figure 4).
\cite{2013ApJ...767L..21W} studied the evolution of the circular velocity function for objects in the Millennium-II simulation \citep{2009MNRAS.398.1150B}, measured at a fixed physical radius of 20 kpc and found "remarkably little evolution since $z=4$" (see their figure 4).
This also indicates that the central region of dark matter halos is already in place since early times (see also our Fig. \ref{fig:halo_buildup}).

The growth of formal virial quantities mainly reflects the underlying cosmic expansion and dilution of matter -- though in a very convoluted way.
The fundamental problem is more that a structural description of a dark matter halo via a single size/mass is not fully adequate.
Within the framework of the spherical collapse model, the concept of virial mass is a perfectly suitable description of the halo.
All the matter within the overdense perturbation ends up in a spherical ball of a fixed size and the rest of the universe is expanding away for eternity.
Therefore, a single size/mass number describes the properties of a halo in this scenario appropriately.
But when we apply this concept to the much more complex configuration of dark matter halos that form in cosmological structure formation simulations, we get the artefact of pseudo-evolution.

The effect of pseudo-evolution is also present for another popular dark matter halo size/mass definition: the friends-of-friends (FoF) method \citep{1985ApJ...292..371D}.
With this method, particles in a simulation are linked together if their mutual distance is smaller than some linking length, which is a free parameter of the method.
Since it is common practice to keep this linking length constant in comoving coordinates, i.e. the linking length grows in physical coordinates, the size and mass definition of FoF halos grows accordingly.
Therefore, the FoF method is also subject to a similar pseudo-evolution effect as in the case for the spherical overdensity size/mass definitions.

Before it was feasible to extract halo tracks from cosmological structure formation simulations, such merger trees were modeled analytically via the extended Press-Schechter formalism \citep{1991ApJ...379..440B,1991MNRAS.248..332B,1993MNRAS.262..627L}.
These analytic prescriptions are still used nowadays since merger trees can be generated much faster, with larger statistics and input parameters (e.g. for a different cosmology) can be varied more easily than for merger trees extracted from simulations.
However, the underlying physical process is still the spherical (or often a more general ellipsoidal) collapse model.
These analytic merger trees are overpredicting the degree of evolution since the mass evolution is very similar to merger trees extracted from simulations where a pseudo-evolving mass definition was used \cite[see e.g.][]{2014MNRAS.440..193J}.

In order to describe the density profiles of dark matter halos, we think it is physically more intuitive to anchor the parametrization on intrinsic properties.
For example in the case of an NFW profile (equation \ref{eq:NFW}), one would specify the scale radius $r_\mathrm{s}$ and normalization $\rho_0$\footnote{Of course, equivalently to using $\rho_0$, one could also specify the mass within a physical scale, e.g. $M(r_\mathrm{s})$ or $M(r_{v_\mathrm{c,max}})$, or fixed physical radius, since these quantities are related with $\rho_0$ via an integral over the density profile.} \citep[see e.g.][]{2008MNRAS.386.1543Z}, instead of linking a formal evolving outer edge $r_\mathrm{vir}$ to a scale radius $r_\mathrm{s}$, which is rather constant, via another evolving parameter, the virial concentration $c_\mathrm{vir}$.
Dark matter halos do not have a well defined outer edge and the density profiles are smooth and can be extended well beyond the formal virial radius \citep{2006ApJ...645.1001P,2008arXiv0807.3027T}.
Thus, anchoring the parametrization of density profiles on intrinsic physical properties seems to be more natural.

Is it then possible to define an edge or a virialized region of a halo with the full phase-space information available in simulations?
The answer is certainly yes.
In Section \ref{sec:virialscale}, we have used the local virial ratio to estimate a virial radius.
\cite{2005MNRAS.363L..11B} and \cite{2008MNRAS.389..385C} use the radial velocity profile to find the equilibrated central region of halos in order to define a size/mass of a halo.
Inspecting the radial velocity patterns around dark matter peaks allows one to identify the static as well as the turnaround region of a halo.
This is probably the most promising method in order to define an associated dynamic mass of a dark matter halo and gives a detailed insight on how the physical mass accretion happens.
In practice, this method can only be done in a satisfactory way by averaging over radial profiles of similar halos since individual halo profiles can be quite noisy.
This is a minor drawback of this method.
\cite{2012ApJ...754..126F} present the novel halo finder ORIGAMI, that tracks the folding of the dark matter sheet in phase-space, which allows for a dynamical definition of a halo.
They also find that their halos are larger than corresponding FoF masses.
All these phase-space based definitions, however, are much more complex to calculate than specifying the mass within a certain radius.
Ultimately, one would like to compare theoretical results from simulations and models with observations, where such definitions are not practical since the necessary phase-space information is not available.
Such definitions might be useful, however, to compare theoretical models with simulation results.

Is it essential to define a virialized region or an outer edge of a halo?
We do not necessarily think so and it might be helpful to liberate ourselves from the Gedankenkorsett\footnote{The literal translation from German is thought-corset.
It describes the limiting reasons why someone cannot think freely or see things from a new perspective, i.e. the box in 'thinking outside the box'.} of the simplistic spherical collapse model.
For example, for the dynamics or gravitational lensing it is not relevant if the matter further out is virialized or not.
The matter is there and acts gravitationally.
This is also reflected in the fact that, for example, other (infalling) halos feel the influence of their future host way beyond the formal virial radius and can be physically affected \citep[e.g.][]{2009MNRAS.398.1742H,2013MNRAS.430.3017B,2014ApJ...787..156B}.
\cite{2008ApJ...680L..25D} have studied the dynamics in the outskirts of a Milky-Way-sized dark matter halo (Via Lactea I) and found that matter (particle) and subhalo orbits typically extend out to about 90 \% of their turnaround radius after their first pericenter passage, i.e. a typical orbit brings subhalos and dark matter particles back to beyond the formal virial radius of a halo on the first two passages \citep[see also][]{2004ogci.conf..513M,2005MNRAS.356.1327G,2009ApJ...692..931L,2013MNRAS.430.3017B}.
Also, the process of virialization is much more complex.
\cite{2007ApJ...667..859D} showed that inner mass shells contract by more than the canonical value of a factor of 2 from the spherical collapse model, whereas outer mass shells contract by a factor smaller than 2 (see their figure 1).
The dynamically affected and collapsing region around a halo is much larger and therefore also the associated mass, which is based on the original definition of the spherical collapse model.
\cite{2011MNRAS.414.3166A} measured this total mass within the turnaround region of a dark matter halo and found that the mass of a halo is significantly higher than conventional virial masses and the mass function is better described by the original \cite{1974ApJ...187..425P} form.

We see that size and mass is a somewhat ill-defined concept in order to describe the complex structure of dark matter halos.
Physical processes (e.g. tidal mass loss or dynamical friction) do not stop or start at a specific outer edge of a halo but their effect strength is rather transitioning smoothly with the strength generally increasing with decreasing distance from the host.
Of course one could argue that a spherical overdensity (or FoF) definition is just a convenient operational definition of halo size/mass, but any such definition is in principle arbitrary and does not indicate a fundamental physical scale.
It seems fine to us to use a spherical overdensity size/mass definition at a \emph{fixed} time or redshift in order to classify or rank the objects as we have done in this study.
However, such simple size/mass definitions do not imply deeper fundamental physical properties like being virialized.
In order to study evolutionary effects, it is not an adequate method though and just obscures the real underlying physical processes.

An additional complication in this matter arises from baryonic physics.
The processes of star formation and feedback can have the potential to affect the dark matter distribution on large scales \citep[e.g.][]{1994ApJ...431..617D,2004ApJ...611L..73K,2009MNRAS.394L..11S,2010MNRAS.407..435A,2010MNRAS.405.2161D,2010ApJ...720L..62K,2010MNRAS.402..776P,2010MNRAS.406..922T, 2012MNRAS.423.2279C,2012ApJ...748...54Z,2013MNRAS.429.3316B}.
Today, the major difficulty in computational structure formation is to simulate these complex processes in a robust way so that one can make reliable predictions.
Despite recent improvements \citep[e.g.][]{2011ApJ...728...88G,2011ApJ...742...76G,2011MNRAS.417..950H,2012MNRAS.427..311W,2012ApJ...748...54Z,2013ApJ...770...25A,2013MNRAS.430.1901H,2013arXiv1311.2073H,2013MNRAS.436.1836R,2013MNRAS.428..129S,2014MNRAS.439.2990S}, this is still an issue \citep[e.g.][]{2012MNRAS.423.1726S} and the improvement and shift from phenomenological to more physical models is the subject of current research in computational galaxy formation.
The changes in the overall dark matter distribution (e.g. contraction or expansion, depending on the details of the modeling) also affect the details of the size or mass definition of a halo \citep[e.g.][]{2009MNRAS.394L..11S,2012MNRAS.423.2279C}.
Also, the baryon fraction is in principle a function of radius \citep[see e.g. figure 5 in][]{2012ApJ...748...54Z} and a single number specified at a pseudo-evolving formal virial radius is not an adequate description of the complex distribution of the baryons.

Of course, this illusion of growth undermines to some degree \emph{any} concept that involves a pseudo-evolving size or mass definition of dark matter halos.
For example, the standard way to describe the abundance of dark matter halos in the universe is to use the mass function, i.e. the mean number of objects of a certain mass per unit volume and unit mass.
The basic functional form of the mass function is inspired by the original \cite{1974ApJ...187..425P} form, but there exist various parametrizations in order to account for different cosmologies, mass definitions and evolutionary trends \citep[e.g.][]{1999MNRAS.308..119S,2001MNRAS.321..372J,2002ApJS..143..241W,2003MNRAS.346..565R,2005Natur.435..629S,2006ApJ...646..881W,2007MNRAS.374....2R,2008ApJ...688..709T,2010MNRAS.402..191P,2011ApJ...732..122B,2011ApJS..195....4M,2013MNRAS.434L..61M}.
The pseudo-evolution of the formal virial mass affects also the classification of halos into host and subhalos, since due to the artificial growth of the virial radius, a neighboring halo could be classified as a subhalo at a later time or for a different size/mass definition.
It has become clear in recent studies that the mass function is not of universal shape when aiming at high precision, i.e. the functional form of the mass function will change as a function of time.
Decoding and understanding then the real physical evolution of the universe from a mass function that uses a pseudo-evolving mass definition is a difficult task.
Another way to describe the abundance is the velocity function, where the maximum circular velocity serves as a 'size' indicator for a halo.
However, the issue of classification into host or subhalo remains in this case as well, since often a formal virial radius is used for this.

In the literature, formal virial quantities also play a major role in describing relations with various other quantities \citep[e.g.][]{2001MNRAS.321..559B,2002ApJ...568...52W,2003MNRAS.339...12Z,2007MNRAS.378...55M,2008MNRAS.391.1940M,2008ApJ...672..122E,2009ApJ...707..354Z,2011MNRAS.413..887C,2013MNRAS.428.3121M} as well as in semi-analytic models of galaxy formation, where such relations are used to encapsulate complex physical processes \citep[e.g.][]{1998MNRAS.295..319M,2000MNRAS.319..168C,2006RPPh...69.3101B,2006MNRAS.365...11C,2010MNRAS.405.1573B,2011MNRAS.413..101G}.
Any correlation between two pseudo-evolving quantities is essentially trivial since the underlying common mechanism is the cosmic expansion.
\cite{2013ApJ...766...25D} illustrate this for the virial concentration - virial mass ($c_\mathrm{vir}-M_\mathrm{vir}$) relation where they show that pseudo-evolution almost entirely accounts for the evolution of this relation at the low-mass end.
It seems difficult to identify plausible processes where a physical quantity should correlate with quantities defined within a pseudo-evolving formal virial region, which just sweeps up matter that was essentially already in place before.
It is potentially more insightful to anchor scaling relations to intrinsic physical quantities that do not show pseudo-evolution.
This makes it much easier to recognize the relevant physical processes.

How should one define the 'size' of a dark matter halo?
There is probably no unique or best definition and this likely depends on the application.
We think the driving force behind this question should be how we compare the results from simulations and models with observations.
It is important for such comparisons that one uses the same definitions so that one is not comparing apples and oranges.
A simple solution in order to avoid pseudo-evolution would be to just keep the threshold density fixed in time as we have done and shown as well in Figs. \ref{fig:mass_evolution} and \ref{fig:size_evolution}.
Any such definition would of course still be arbitrary and not define a fundamental physical scale of a dark matter halo but in this way one keeps the "meter-stick" constant and one can better disentangle real physical growth from pseudo-evolution.
This should work fine in recent times and we have done this in this study for the last 10 Gyr, i.e. up to ca. $z \approx 1.6$.
However, this method is doomed to fail at early cosmic times, depending on the exact threshold density used.
For example, in the case of keeping the threshold density at the value of 200 times the mean matter density today (200m), the mean matter density of the universe has reached that value at the redshift of $z = 200^{1/3} - 1 = 4.848$.
Therefore, one cannot use the same definition in order to describe halos at high redshift and today.
In practice, however, the time or redshift range of interest is often not that extreme (e.g. limited redshift range in an observational survey) and a definition with a fixed density threshold in time is potentially a good way to avoid the effect of pseudo-evolution.
This would also be convenient in a practical sense, since one can use existing software tools with only modest modifications.
Also intrinsic physical scales like the maximum value of the circular velocity $v_\mathrm{c,max}$ (and $r_{v_\mathrm{c,max}}$), the concentration $c_\mathrm{v}$ or the scale radius $r_\mathrm{s}$, which is physically increasing with halo size/mass, could be used to describe the 'size' of a dark matter halo and are potentially also easier to compare directly with observations.
The usefulness of $v_\mathrm{c,max}$ and $c_\mathrm{v}$ as halo scales has been pointed out before and they are also better suited to describe subhalos \cite[e.g.][]{1999ApJ...516..530K,2007ApJ...667..859D,2008MNRAS.386.2022A,2011MNRAS.415.2293K,2013MNRAS.435.1618K}.

\begin{figure}
  \begin{center}
    \includegraphics[width=\columnwidth]{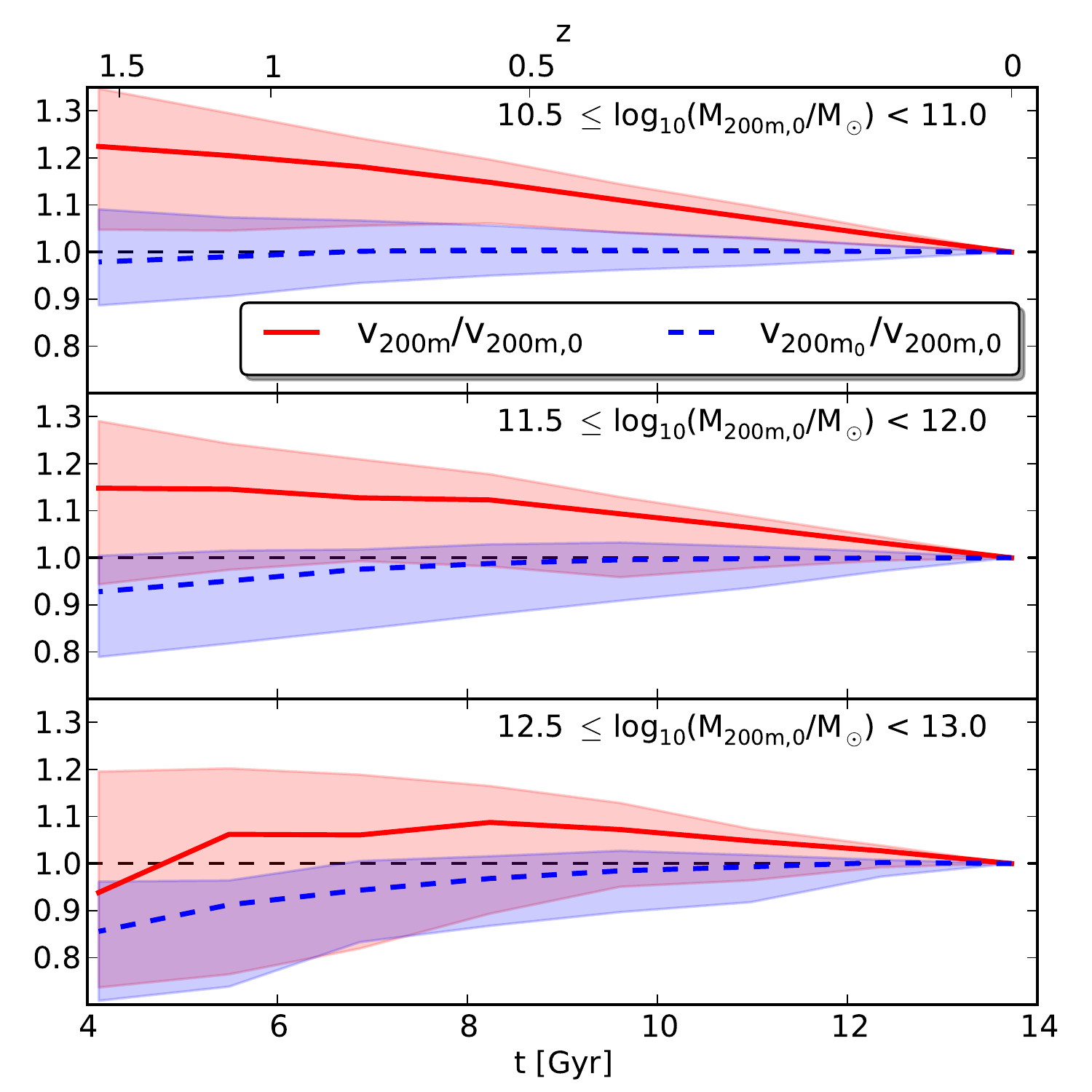}
    \caption{Relative evolution of the formal virial velocity $v_\mathrm{200m}$ over the last 10 Gyr compared with the evolution of the circular velocity $v_\mathrm{200m_0}$.
    Plotted are the medians over all halo tracks (lines) as well as the range between the 15th and 85th percentile (shaded area).
    The pseudo-evolution of the virial velocity $v_\mathrm{200m}$ is rather moderate since the individual pseudo-evolution of size and mass cancel to some degree.
    Thus, this is rather a property of NFW-like density profiles than a specific fundamental property at a formal virial radius.}
    \label{fig:virial_velocity_evolution}
  \end{center}
\end{figure}

In general, a more holistic approach and look at the overall radial profile of physical quantities and a parametrization based on intrinsic physical scales is potentially a promising way in order to better describe and understand the structure and evolution dark matter halos.
For example, another important (derived) virial scale used in the literature is the virial velocity, $v_\mathrm{vir} \equiv \sqrt{G M_\mathrm{vir}/r_\mathrm{vir}}$.
This is just the value of the circular velocity at the formal virial radius.
The virial velocity is often used in order to set a characteristic energy scale (e.g. $T_\mathrm{vir} \propto v_\mathrm{vir}^2$) of the halo. 

In Fig. \ref{fig:virial_velocity_evolution}, we see that the formal virial velocity $v_\mathrm{200m}$ is also pseudo-evolving when compared to the value of $v_\mathrm{200m_0}$ (fixed background density at $z=0$).
The pseudo-evolution of $v_\mathrm{200m}$ is rather small when compared to the effect for the size or mass of a halo as we have seen in previous sections (see e.g. Figs. \ref{fig:mass_evolution} and \ref{fig:size_evolution}).
Such a modest evolution of the virial velocity is also seen in \cite{2013ApJ...767L..21W} for a large sample of dark matter halos from the Millennium-II simulation (see their figure 4).
This is due to the partial cancellation of the pseudo-evolution of size and mass for the circular velocity $v_\mathrm{c}(r) \equiv \sqrt{G M(r)/r}$.
Differently viewed, it is rather a characteristics of the shape of the circular velocity curve for a density profile similar to the NFW shape that exhibits a flattish plateau after the peak is reached than a fundamental scale at the formal virial radius.

Similarly, in \cite{2013ApJ...779..159D} they investigate the relation between mass and velocity dispersion ($M$-$\sigma$).
They show that this relation for clusters not only holds at a formal virial scale but over a wide radial range.
Thus it is the result of tight equilibrium relations between radial profiles of physical quantities, rather than a fundamental property due to the virial theorem.
This is also a nice illustration, that the apparent fundamental relation at a formal virial scale was obscuring a more general relation between physical quantities that holds over a wider radial range.
There are likely many more such cases.

\begin{figure}
  \begin{center}
    \includegraphics[width=\columnwidth]{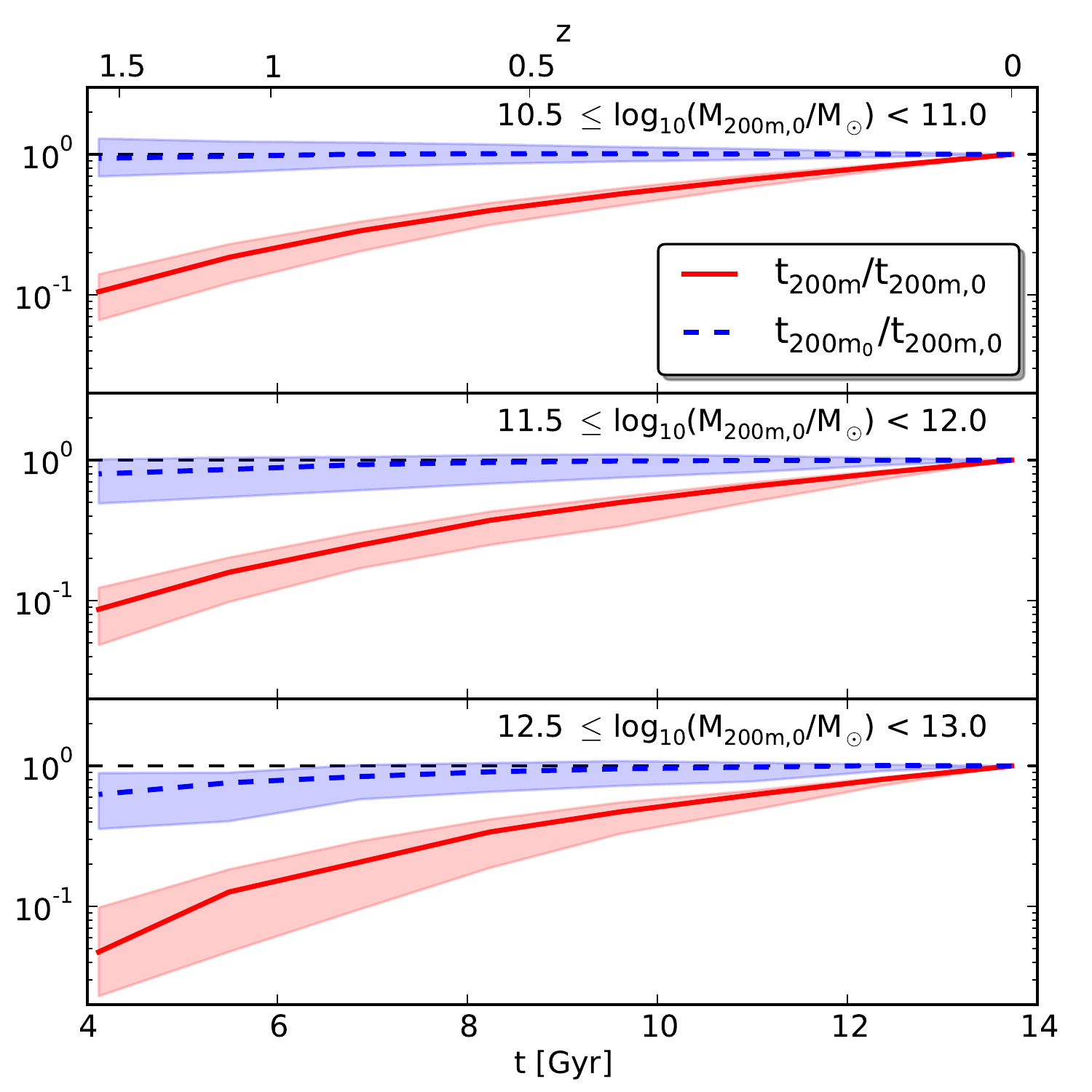}
    \caption{Relative evolution of the dynamical time at the formal virial radius, $t_\mathrm{200m}$, over the last 10 Gyr.
    Plotted are the medians over all halo tracks (lines) as well as the range between the 15th and 85th percentile (shaded area).
    We compare this with the evolution of the dynamical time at $r_\mathrm{200m_0}$, which better reflects the actual physical evolution of the dark matter halo.
    It is evident that $t_\mathrm{200m}$ gives the impression of a small characteristic time scale of the dark matter halo at high redshift.}
    \label{fig:virial_time_evolution}
  \end{center}
\end{figure}

Another derived virial scale is the dynamical time at the formal virial radius, $t_\mathrm{vir} \equiv \sqrt{r_\mathrm{vir}^{3}/G M_\mathrm{vir}} = r_\mathrm{vir}/v_\mathrm{vir}$, which often serves as a characteristic time scale for physical processes (e.g. cooling or feedback) in analytic models.
In Fig. \ref{fig:virial_time_evolution}, we compare the evolution of the time scale $t_\mathrm{200m}$ with the time scale $t_\mathrm{200m_0}$.
We can clearly see that for no physical reason the characteristic time $t_\mathrm{200m}$ would become much shorter at early times than a more physically motivated time scale $t_\mathrm{200m_0}$.
Thus, the general discrepancy between physical scales and formal virial scales could potentially account for some fundamental problems in (semi-)analytic models like, e.g., low-mass galaxy evolution \citep{2012MNRAS.426.2797W}.

The current status quo of how we describe the structure and evolution of dark matter halos in our universe is somewhat unsatisfactory since often the underlying physical processes are obscured by an unfortunate, pseudo-evolving parametrization or description.
The fundamental problem here is, that simplistic analytic models of structure formation cannot capture the complex processes seen in cosmological structure formation simulations.
Using then concepts based on formal virial quantities can give a distorted picture that does not properly reflect the actual physical evolution of dark matter halos.

\section{Summary}\label{sec:summary}

In this study, we have critically investigated the pseudo-evolution effect of sizes and masses of dark matter halos due to a spherical overdensity definition.
We summarize our findings as follows:
\begin{itemize}
  \item The commonly used method of defining sizes and masses of dark matter halos based on an evolving overdensity threshold shows a large degree of pseudo-evolution when compared to the size or mass evolution within fixed physical scales or the evolution of intrinsic physical properties (see Figs. \ref{fig:mass_evolution}, \ref{fig:size_evolution}, \ref{fig:vcmax_evolution}, \ref{fig:rs_evolution} and \ref{fig:halo_buildup}).
  \item The pseudo-evolution is more pronounced for low-mass objects, which essentially have their density profiles in place 10 Gyr ago even beyond a formal virial radius.
    However, also for group and cluster sized objects, which are still physically accreting at $z=0$, the pseudo-evolution is substantial.
    The degree of pseudo-evolution also depends on how much the threshold density varies (see Figs. \ref{fig:rhoSO} and \ref{fig:mass_evolution_relative}).
  \item The FoF size and mass definitions are subject to a similar pseudo-evolution effect as the spherical overdensity definitions.
  \item Analytic merger trees are overestimating the mass growth since they agree well with merger trees extracted from cosmological structure formation simulations that use a pseudo-evolving mass definition.
  This indicates that it is difficult to capture the complex mass assembly of dark matter halos with simplistic models.
  \item The concept of size or edge of a dark matter halo is somewhat ill-defined and one can in practice only specify operationally defined scales.
    Such 'size' definitions are in principle arbitrary and do in general not coincide with the virialized region.
	\item It is a physically more appropriate nomenclature to name these halo scales by their characteristics, e.g. $r_\mathrm{200m}$/$M_\mathrm{200m}$ or $r_\mathrm{180c}$/$M_\mathrm{180c}$, and one should refrain from calling them virial radius or virial mass, since this is a deeper physical property that is not implied by such simple 'size'  definitions.
  \item In general, we question the widely accepted view of attributing pseudo-evolving formal virial quantities central importance for describing the structure and evolution of dark matter halos (and sometimes even of galaxies within the halos, as done in semi-analytic models).
		Concepts and relations based on pseudo-evolving formal virial quantities do not properly represent the actual evolution of dark matter halos and lead to an inaccurate picture of the physical evolution of our universe.
    In order to gain a better physical picture of our universe, it might be auspicious to parametrize such relations with respect to intrinsic physical scales of dark matter halos or with 'size' measures that do not show pseudo-evolution.
\end{itemize}

\acknowledgments

It is a great pleasure to thank J\"{u}rg Diemand for many discussions during our common time in Zurich and Santa Cruz, where many issues presented in this study were debated over coffee.
We also thank J\"{u}rg Diemand, Benedikt Diemer, Andrey Kravtsov and Amber Kilgore for comments on a draft version of this paper.
We would also like to thank Yinghe Lu for interesting questions and discussions during her Bachelor thesis in the academic year 2012/2013 at Peking University, where she was working on aspects of this work.
MZ is supported by a 985 grant from Peking University and the International Young Scientist grant 11250110052 by the National Science Foundation of China.
The simulation in this work has been performed on the Flux cluster at the University of Michigan.
This research has made use of NASA's Astrophysics Data System (ADS), the arXiv.org preprint server, and the Python plotting library Matplotlib \citep{2007CSE.....9...90H}.

\appendix

\section{Pseudo-evolution mechanism}\label{sec:mechanism}

\begin{figure}
  \begin{center}
    \includegraphics[width=0.5\columnwidth]{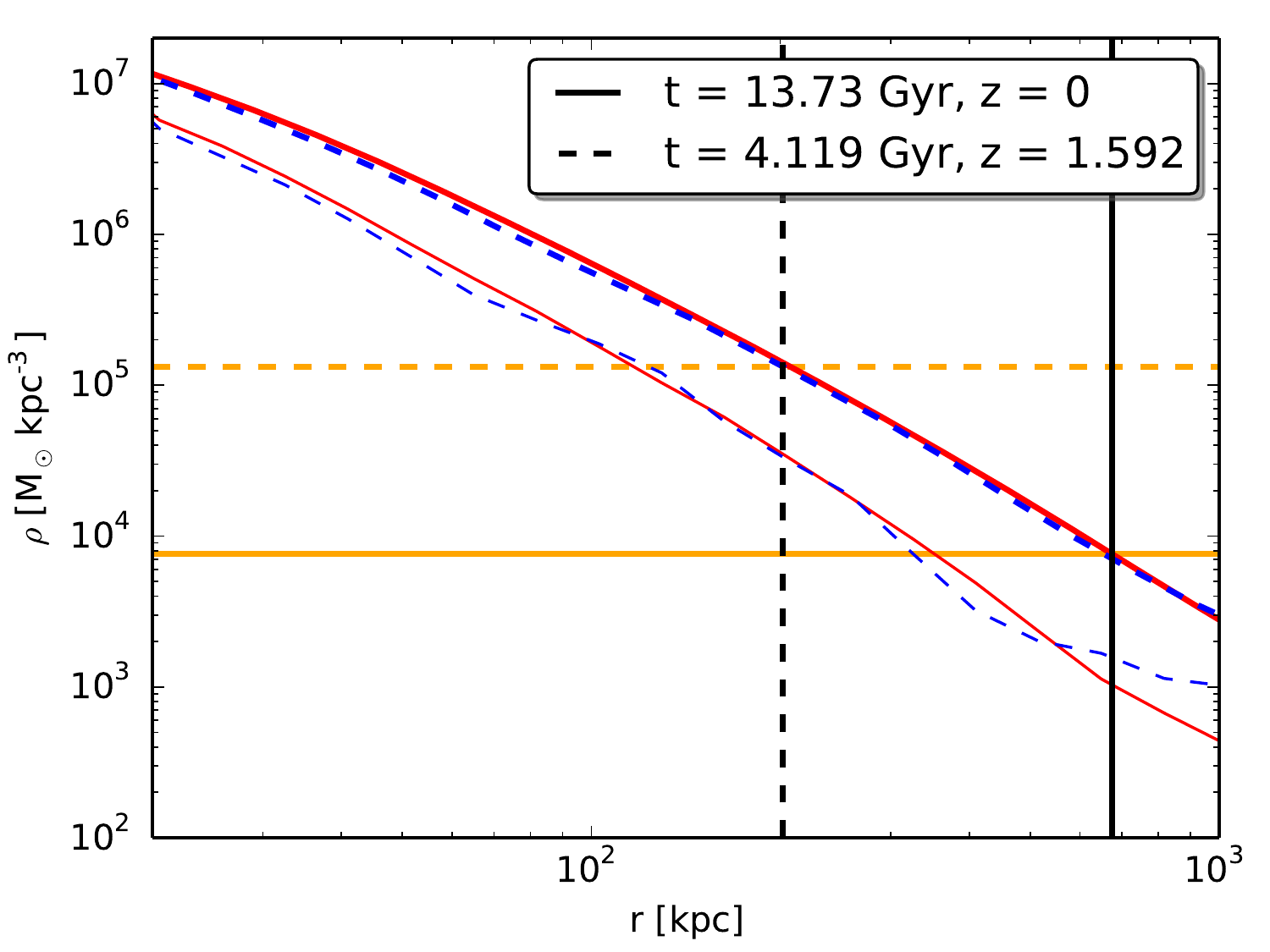}
    \caption{Illustration of the pseudo-evolution mechanism for an individual dark matter halo.
    For all quantities shown, the dashed lines show the values at $z=1.592$ and the solid lines the values today at $z=0$, i.e. a time span of ca. 10 Gyr.
    The horizontal lines show that the value for $200 \rho_\mathrm{mean}$ has dropped by a factor of 17.42 between those 2 snapshots.
    The thin diagonal lines denote the density profile $\rho(r)$ and the normal diagonal lines denote the enclosed density profile $\rho_\mathrm{enc}(r) \equiv M(r)/ (4 \pi r^{3}/3)$ in physical coordinates.
    From the density profiles it is clearly evident, that there is not much actual physical evolution for this object in the last 10 Gyr.
    This is also reflected in the rather moderate change of the peak circular velocity value $v_\mathrm{c,max}$ from $321.8 \, \mathrm{km} \, \mathrm{s}^{-1}$ to $342.4 \, \mathrm{km} \, \mathrm{s}^{-1}$ (ca. 6\% increase) during that time interval.
    However, the 200m size/mass definition would suggest that the formal virial radius $r_\mathrm{200m}$ (vertical lines) has grown from 202.0 kpc to 674.6 kpc (ca. a factor 3.3 increase) and the virial mass $M_\mathrm{200m}$ from $4.580 \times 10^{12} \, \Mo$ to $9.793 \times 10^{12} \, \Mo$ (ca. a factor 2.1 increase).
    }
    \label{fig:pseudo_evolution}
  \end{center}
\end{figure}

In Fig. \ref{fig:pseudo_evolution}, we illustrate the pseudo-evolution mechanism for an individual dark matter halo.
For this object, the 200m size/mass definition would suggest that the formal virial radius $r_\mathrm{200m}$ has grown from 202.0 kpc to 674.6 kpc (ca. a factor 3.3 increase) and the virial mass $M_\mathrm{200m}$ from $4.580 \times 10^{12} \, \Mo$ to $9.793 \times 10^{12} \, \Mo$ (ca. a factor 2.1 increase) in the last 10 Gyr.
However, the density profiles do not show much physical evolution in the same time interval which is also reflected in the only 6\% increase in the peak circular velocity value.
At redshift $z=1.592$, the mass within the formal virial radius at $z=0$ was $M(r_\mathrm{200m,0}) = 9.130 \times 10^{12} \, \Mo$.
This means that 93\% of the mass within the physical scale $r_\mathrm{200m,0}$ was already present 10 Gyr ago for this dark matter halo.

\section{Evolution of mass ratios}\label{sec:massratio}

\begin{figure}
  \begin{center}
    \includegraphics[width=0.5\columnwidth]{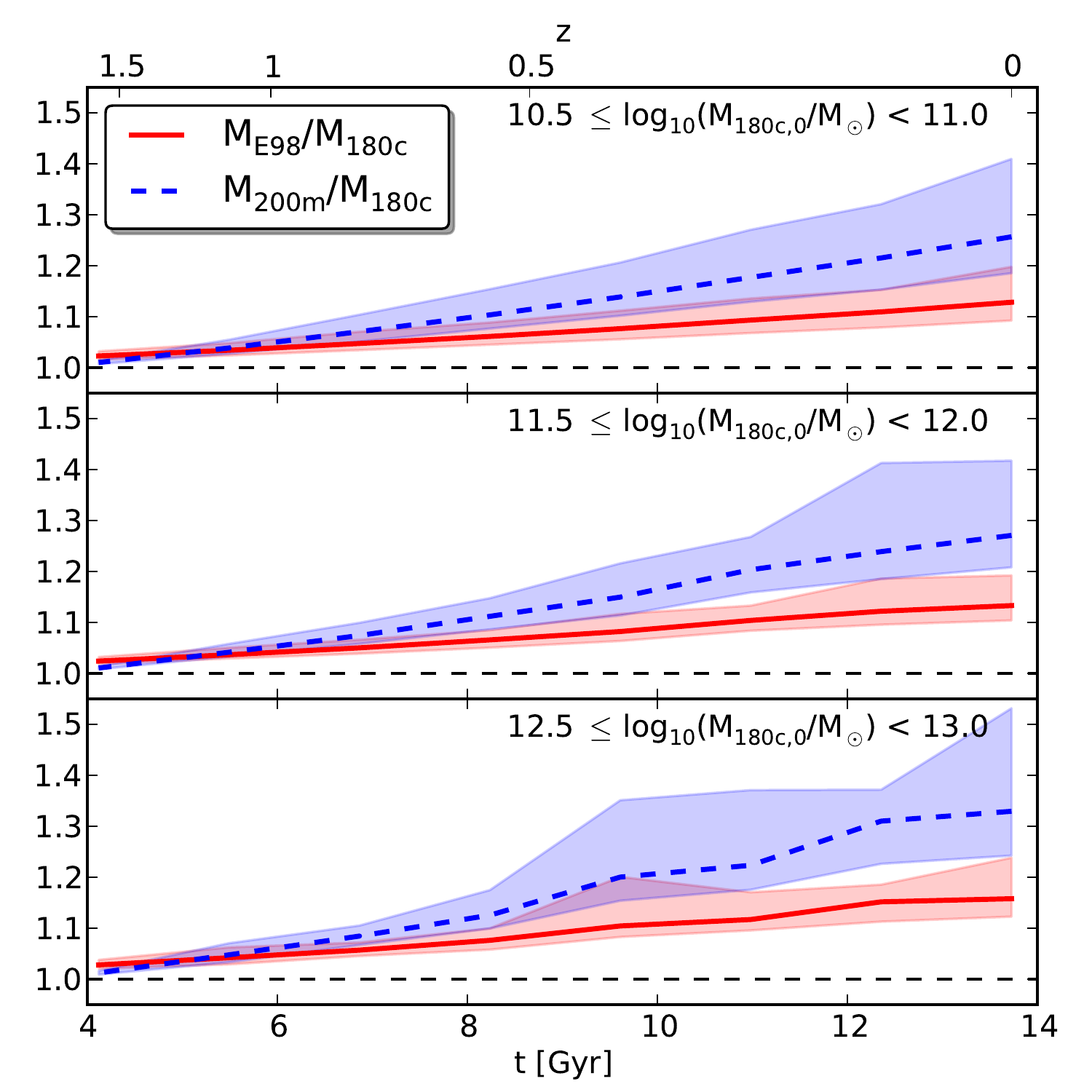}
    \caption{Evolution of the mass ratios $M_\mathrm{E98}/M_\mathrm{180c}$ and $M_\mathrm{200m}/M_\mathrm{180c}$ over the last 10 Gyr.
    Plotted are the medians over the halo tracks in the specific mass class as well as the range between the 15th and 85th percentile as shaded region.
    The mass ratios grow with time as well as with halo mass.}
    \label{fig:mass_ratio_evolution}
  \end{center}
\end{figure}

Here, we illustrate the evolution of the mass ratio of the 3 size/mass definitions used in this work.
In Fig. \ref{fig:mass_ratio_evolution}, we plot $M_\mathrm{E98}/M_\mathrm{180c}$ and $M_\mathrm{200m}/M_\mathrm{180c}$ over the last 10 Gyr for 3 mass categories.
In general, the discrepancy between the various definitions becomes larger with time as can be intuitively understood from the evolution of the defining threshold density in Fig. \ref{fig:rhoSO}.
The ratios do not only grow with time but become also larger for more massive objects.
Therefore, there can be quite a difference in halo mass for the same object, depending on the used definition \citep[see also][]{2001A&A...367...27W,2003ApJ...584..702H}.
The exact definition of the size of a halo also influences the classification of objects as host or subhalos, which in turn affects the abundance (i.e. mass function) of the dark matter halos.

\section{Estimation of the virial ratio}\label{sec:virialdetails}

In each spherical shell, we calculate the velocities of the dark matter particles with respect to the halo center.
We convert them to physical velocities by adding the coordinate expansion (Hubble flow) since the simulation was performed in comoving coordinates.
The kinetic term in the shell is then simply calculated by
\begin{eqnarray}
K &\equiv& \frac{1}{2} \sum_{i} m_i \left(v_\mathrm{rad,i}^{2} + v_\mathrm{\varphi,i}^{2} + v_\mathrm{\theta,i}^{2} \right) \\ 
&=& \frac{1}{2} M \left(\bar{v}_\mathrm{rad}^{2} + \bar{v}_\mathrm{\varphi}^{2} + \bar{v}_\mathrm{\theta}^{2} + \sigma_\mathrm{rad}^{2} + \sigma_\mathrm{\varphi}^{2} + \sigma_\mathrm{\theta}^{2} \right)
\end{eqnarray}
where $M = \sum_{i} m_i$ is the total mass in the shell and the sum is over all particles in the shell.
We have decomposed the velocity field in spherical coordinates.
The second form allows us to use the mean velocities and dispersions for the different velocity components calculated by the profiling tool.

For an isolated, spherically symmetric structure, the potential at $r$ is given by \cite[equation 2.28]{2008gady.book.....B}
\begin{equation}\label{eq:pot}
  \Phi(r) = -\frac{G M(r)}{r} - G \int_{r}^{\infty} \frac{\mathrm{d} M(r^\prime)}{r^\prime} \, .
\end{equation}
In cosmology, the potential results from the density perturbations and we can remove the contribution of the constant background density \citep[Jeans Swindle, see e.g.][]{2008gady.book.....B,2013MNRAS.431L...6F}.
In practice, we only do the integration of the outer part of the potential out to a finite radius instead of infinity or until we reach the mean matter density.
We have checked and found it not to be very sensitive on the exact value as long as it is of the order of a few Mpc, as in our case.
We then interpolate on the resulting potential to get the value at the middle radius of each shell.
The potential energy of the mass M in each shell with middle radius $r_\mathrm{m}$ is then simply given by
\begin{equation}
  W = \frac{1}{2} M \Phi(r_\mathrm{m}) \, .
\end{equation}

Here, a side remark is in place.
Often one finds studies that use the local (differential), as in this study, or cumulative version of the virial ratio $Q$ as a function of radius, where it shows an upturn towards the center of the halo with large values for $Q$ that would mean the matter in the center is unbound, i.e. $Q > 2$ \citep[e.g.][]{1996MNRAS.281..716C,2003ApJ...588...35M,2006ApJ...646..815S}.
Of course, this does not make physical sense and is an artifact of neglecting the contribution of the outer part of the potential (second term on the right-hand side in equation (\ref{eq:pot})) as we have verified.
In \cite{2006ApJ...646..815S} they remedy this situation by including a surface pressure term.

\bibliography{RDB_S}

\end{document}